\documentclass[12pt]{iopart}

\usepackage{epsf}
\usepackage{graphicx}

\begin{document}

\def\nuc#1#2{${}^{#1}$#2}
\def\mee{$\langle m_{\beta\beta} \rangle$}
\def\mnu{$\langle m_{\nu} \rangle$}
\def\gnu{$\langle g_{\nu,\chi}\rangle$}
\def\mmod{$\| \langle m_{\beta\beta} \rangle \|$}
\def\mb{$\langle m_{\beta} \rangle$}
\def\BBz{$\beta\beta(0\nu)$}
\def\BBm{$\beta\beta(0\nu,\chi)$}
\def\BBt{$\beta\beta(2\nu)$}
\def\BB{$\beta\beta$}
\def\Mz{$|M_{0\nu}|$}
\def\Mt{$|M_{2\nu}|$}
\def\Tz{$T^{0\nu}_{1/2}$}
\def\Tt{$T^{2\nu}_{1/2}$}
\def\Tc{$T^{0\nu\,\chi}_{1/2}$}
\def\ms{$\delta m_{\rm sol}^{2}$}
\def\ma{$\delta m_{\rm atm}^{2}$}
\def\ts{$\theta_{\rm sol}$}
\def\ta{$\theta_{\rm atm}$}
\def\tot{$\theta_{13}$}
\def\be{\begin{equation}}
\def\ee{\end{equation}}
\def\today{\space\number\day\space\ifcase\month\or January\or February\or
    March\or April\or May\or June\or July\or August\or September\or October\or
    November\or December\fi\space\number\year}

\newcommand{\simgt}{\ \raisebox{-.25ex}{$\stackrel{>}{\scriptstyle \sim}$}\ }
\newcommand{\simlt}{\ \raisebox{-.25ex}{$\stackrel{<}{\scriptstyle \sim}$}\ }

\title[Double Beta Decay]{Double Beta Decay}

\author{Steven R. Elliott$\dag$, and Jonathan Engel$\ddag$ }

\address{\dag\ Los Alamos National Laboratory, MS H803, Los Alamos,
NM 87545, USA}

\address{\ddag\ University of North Carolina, Department of Physics
and Astronomy, Chapel Hill, NC 27599-3255, USA}

\begin{abstract}
We review recent developments in double-beta decay, focusing on what 
can be learned
about the three light neutrinos in future experiments.  We examine 
the effects of
uncertainties in already measured neutrino parameters and in calculated nuclear
matrix elements on the interpretation of upcoming double-beta decay 
measurements.
We then review a number of proposed experiments.

\end{abstract}


\tableofcontents


\maketitle

\section{Introduction}
Double-beta (\BB) decay is a second-order weak process in which two neutrons
inside a nucleus spontaneously transform into two protons.  To conserve charge
two electrons must be emitted.  If lepton number is also to be conserved two
antineutrinos must be emitted as well.  This lepton-number-conserving
process --- \BBt\ decay
--- has been observed in several nuclei.  Lepton number is not
associated with a
gauge symmetry, however, and so its conservation is not sacrosanct. If lepton
number is violated, e.g.\ through the propagation of Majorana neutrinos, then a
variant of the decay in which no neutrinos are emitted --- \BBz\ decay
--- may also occur, though it has never been observed. In \BBz\ decay a virtual
neutrino is emitted by one neutron and absorbed by the second. \BBz\ decay can
also occur through the exchange of other particles, perhaps some predicted by
supersymmetric models, between the nucleons.  Although we address this and
other more exotic possibilities here, we are most interested in \BBz\ decay
mediated by light Majorana neutrinos because its rate depends on the absolute
neutrino mass scale, a number on which we currently have only a generous upper
limit.

In recent years experimenters have discovered that neutrinos have nonzero
masses and mixings, and have pinned down many of the associated parameters. But
although we now know the differences between the squares of the masses, we do
not know the mass of the lightest neutrino, nor the pattern in which masses of
the three active neutrinos are arranged.  \BBz\ experiments have the potential
to teach us about these matters, and in this review we focus on the question of
how much we can expect to learn from experiments in the next decade.

Previous reviews have thoroughly addressed other aspects of \BB\ decay. Early
papers by Primakoff and Rosen (1959), Haxton and Stephenson (1984),
and Doi \etal (1985), as well as
more recent reviews by Tomoda (1991), Suhonen and Civitarese (1998),
Vergados (2000)
and Klapdor-Kleingrothaus (2000) presented
the theoretical formalism in great detail.  A comprehensive review by
Faessler and \v{S}imkovic (1998) in this journal  devoted particular
attention to \BBz\ decay in supersymmetric models and to the calculation of the
nuclear matrix elements governing all kinds of \BB\ decay in the Quasiparticle
Random Phase Approximation (QRPA).  We will treat these topics,
particularly the first, more briefly. A recent
review by Elliott and Vogel (2002) examined the impact of
recent discoveries in neutrino physics on \BBz\ decay, but since then several
experiments (e.g.\ KamLAND and WMAP) have reported results and a number of
papers interpreting the results have appeared, with the consequence that
neutrino masses and mixings are better understood. We will focus most intently
on the \textit{additional} neutrino physics that can be extracted from \BBz\
decay in light of these very recent results. The discussion will require us to
examine the accuracy with which the \BBz\ nuclear matrix elements can be
calculated, in addition to surveying present and future experiments.

\section{General Theory of \BB\ Decay }
This subject has been covered extensively in the reviews listed above and we do
not present it in great detail here.  Since we focus on what can be learned
about neutrinos from \BBz\ decay, we must begin with a brief discussion of
Majorana particles.

We define the left and right handed components of a Dirac 4-spinor by
$\Psi_{L,R}=[(1\mp\gamma_5)/2 ]\Psi$.  In the Standard Model, only left-handed
neutrinos interact.  Because neutrinos are neutral, however, there is an
additional way to construct left-handed neutrino fields.  The charge
conjugate field $\Psi^c \equiv \rmi \gamma^2 \Psi^*$ is also
neutral; neutrino fields can be linear combinations of $\Psi$
and $\Psi^c$ and, since $(\Psi_R)^c$ is left-handed (and $(\Psi_L)^c$
is right-handed) we can
define independent Majorana neutrino fields that are their own charge
conjugates (antiparticles) via
\begin{equation}
\nu=\frac{\Psi_L+(\Psi_L)^c}{\sqrt{2}}~, ~~~
X=\frac{\Psi_R+(\Psi_R)^c}{\sqrt{2}}~.
\end{equation}
Mass terms in the Lagrangian can couple these two kinds
of fields to themselves and each other; the most general mass term
has the form
\begin{equation}
\mathcal{L}_M= -M_L \bar{\nu} \nu - M_R \bar{X} X - M_D (\bar{\nu} X + \bar{X}
\nu)~.
\end{equation}
The parameters $M_L$ and $M_R$ are ``Majorana" masses for the $\nu$ and $X$
fields, and $M_D$ is a ``Dirac" mass that couples the two.  For $N$ flavors of
neutrinos these masses can be arranged in a matrix:
\begin{equation}
\mathcal{L}_M= -(\bar{\nu}~ \bar{X}) \mathcal{M} \left(\begin{array}{c}
    \nu\\X \end{array} \right) ~~~\mathcal{M}=\left( \begin{array}{cc}
\mathcal{M}_L~ \mathcal{M}_D^T\\ \mathcal{M}_D~ \mathcal{M}_R
\end{array}
\right)
~,
\end{equation}
where the
matrices $\mathcal{M}_L$, $\mathcal{M}_R$, and $\mathcal{M}_D$ are now $N
\times N$.  If  $\mathcal{M}_L$ and $\mathcal{M}_R$ are zero, the
Majorana $\nu$'s
pair with an $X$ to form $N$ Dirac neutrinos.  At another extreme, in
the
``see-saw''
mechanism (Gell-Mann {\it et al}  1979; Yanagida, 1979; Mohapatra and
Senjanovic  1980) --- here with one flavor, for
simplicity --- $M_L << M_D << M_R$ and
the resulting eigenstates are Majorana neutrinos, the lightest of
which has mass $m \sim M_D^2/M_R$.

In general the eigenstates of $\mathcal{M}$ represent Majorana
neutrinos that are related to
the
fields $\nu$ and $X$ by the ``mixing''
matrix of eigenvectors:
\begin{equation}
  \left(\begin{array}{c} \nu\\X  \end{array}\right) =
  \left(\begin{array}{c} U\\V  \end{array} \right)
  \left(\begin{array}{c} \phi\\\Phi  \end{array}\right)~,
\end{equation}
where $U$ and $V$ are $N \times 2N$ submatrices.
We assume here that the states $\Phi$ are either
absent or extremely heavy, in which case we can work with a nearly
unitary  $N \times N$ section of
$U$ that mixes the light neutrinos:
\begin{equation}
\nu_l \simeq \sum_m U_{lm} \phi_m~.
\end{equation}
The original states $\nu_l$ with definite flavor that enter the
Lagrangian are linear combinations of the states $\phi_m$ with definite mass.

The matrix $U$ nominally has $N^2$ parameters, $N(N-1)/2$ angles and
$N(N+1)/2$ phases.   $N$ of the phases are unphysical (Kobazarev \etal  1980),
leaving
$N(N-1)/2$ independent physical phases.  For three active neutrinos, the mixing
matrix can be written in the form
\begin{equation}
\fl U = \left(\begin{array}{ccc} c_{12} c_{13} & s_{12} c_{13} & s_{13} e^{-i
\delta}
  \\
-s_{12} c_{23} - c_{12} s_{23} s_{13} e^{i \delta} & c_{12} c_{23} - s_{12}
s_{23} s_{13} e^{ i \delta} & s_{23} c_{13}
\\
s_{12} s_{23} - c_{12} c_{23} s_{13} e^{i \delta} & -c_{12} s_{23} - s_{12}
c_{23} s_{13} \rme^{\rmi \delta} & c_{23} c_{13}
\end{array}\right) 
\mathrm{diag}\{e^{\frac{i\alpha_1}{2}},e^{\frac{i\alpha_2}{2}},1\}~~,
\end{equation}
where $s_{ij}$ and $c_{ij}$ stand for the sine and cosine of the
angles $\theta_{ij}$, $\delta$ is a ``Dirac" phase analogous to
the phase in the CKM matrix, and the other two phases $\alpha_1$ and $\alpha_2$
affect only Majorana particles. (They can't be rotated away because Majorana
neutrinos don't conserve lepton number.)

The neutrino masses and mixing matrix figure in the rate of neutrino-mediated
\BBz\ decay. If the weak quark-lepton effective low-energy Lagrangian has the
usual V-A form, then the rate for that process is
\begin{equation}
\label{eq:rate} [T^{0\nu}_{1/2}]^{-1}=\sum_{\textrm{spins}} \int |Z_{0\nu}|^2
\delta(E_{e1}+E_{e2}-\Delta E) \frac{\rmd^3p_1}{2\pi^3}\frac{\rmd^3p_2}
{2\pi^3}~,
\end{equation}
where $Z_{0\nu}$ is the amplitude for the process and $\Delta E$ is the Q-value
of the decay.

The amplitude $Z_{0\nu}$ is evaluated in second-order perturbation theory and
can be written as a lepton part contracted with a hadron part. The lepton part
of the amplitude,
containing outgoing electrons and exchanged virtual Majorana neutrinos of mass
$m_j$, emitted and absorbed with amplitude $U_{ej}$, is
\begin{equation}
\label{eq:lep}
-\frac{i}{4} \int \sum_j \frac{\rmd^4q}{(2\pi)^4}\rme^{-\rmi q\cdot(x-y)}
\bar{e}(x) \gamma_{\mu}(1-\gamma_5) \frac{q^{\rho}\gamma_{\rho}+m_j}{q^2-m_j^2}
(1-\gamma_5) \gamma_{\nu} e^c(y)~U_{ej}^2~,
\end{equation}
where $q$ is the 4-momentum transfer.  The term with $q^{\rho}$
vanishes and the $m_j$ in the denominator can be neglected for light
neutrinos, so that the amplitude is proportional to
  \begin{equation}
   \label{eq:mass}
    \langle m_{\beta\beta} \rangle = \left| \sum_j m_j U^2_{ej}\right| 
= \left| m_1
    |U_{e1}|^2 + m_2
    |U_{e2}|^2 \rme^{\rmi(\alpha_2-\alpha_1)} + |U_{e3}|^2 \rme^{\rmi
    (-\alpha_1-2\delta)}
    \right|~.
  \end{equation}
The absolute value has been inserted for convenience, since the quantity 
inside
it is squared in equation (\ref{eq:rate}) and is complex if CP is violated.

The hadronic part of the amplitude must be evaluated between initial and final
nuclear ground states, with the states in the intermediate nucleus summed over
in the same way the virtual neutrino's momentum must be integrated over.  For
the amplitude to be appreciable, the wavelength of the virtual neutrino cannot
be more than a few times larger than the nuclear radius $R$, i.e.\ only momenta
$q \simgt 1/R \sim 100$ MeV will contribute.  The excitation energies of the
intermediate nuclear states generated by the hadronic part of the amplitude are
all significantly smaller and so to good approximation the individual energies
of those states in the ``energy denominator" can be neglected or replaced by an
average value $\bar{E}$ (to which the expression is not very sensitive) and the
states can be summed over in closure. The result, in the allowed approximation
for the weak hadronic current, is

\begin{eqnarray}
\label{eq:0nurate}
   [T^{0\nu}_{1/2}]^{-1} & = & G_{0\nu}(\Delta E,Z)
  \left|M^{GT}_{0\nu}-\frac{g_V^2}{g_A^2}
  M^{F}_{0\nu}\right|^2 \langle m_{\beta\beta} \rangle^2  \nonumber \\
& \equiv & G_{0\nu}(\Delta E,Z)
  \left|M_{0\nu}\right|^2 \langle m_{\beta\beta} \rangle^2~.
\end{eqnarray}

Here $G_{0\nu}(\Delta E,Z)$ comes from the phase-space integral (which
depends on the nuclear charge $Z$ through the wave functions of the
outgoing electrons),
$g_A$ and $g_V$ are the weak axial-vector and vector coupling
constants, and the $M$'s, which are nuclear matrix elements of
Gamow-Teller-like and Fermi-like two-body operators, are defined as
\begin{eqnarray}
   M^{F}_{0\nu}&=& \langle f |\sum_{j,k} H(r_{jk},\bar{E}) \tau^+_j \tau^+_k
   |i\rangle \\
   M^{GT}_{0\nu}&=& \langle f |\sum_{j,k} H(r_{jk},\bar{E}) \vec{\sigma}_j 
\cdot \vec{\sigma}_k \tau^+_j \tau^+_k |i\rangle ~.
\end{eqnarray}
The function $H$, which depends on the distance between nucleons and
(quite weakly)
on the average nuclear excitation energy in the intermediate nucleus, is
sometimes called a ``neutrino potential" and has approximate form
\begin{equation}
   H(r,\bar{E})=\frac{2R}{\pi r} \int_0^{\infty} \rmd q \frac{q \sin{qr}}
{\omega
   (\omega + \bar{E}-[M_i+M_f]/2)}~,
\end{equation}
where $M_i$ and $M_f$ are the masses of the initial and final nuclei.

The approximate expression equation (\ref{eq:0nurate}) is typically accurate to
within about 30\%, the largest corrections coming from ``induced" terms (weak
magnetism, induced pseudoscalar) in the hadronic current (\v{S}imkovic
{\it et al}  1999). For \BBt\ decay, the
expression for the rate [\Tt]$^{-1}$ is similar to equation
(\ref{eq:0nurate}), the differences being in the phase space factor
$G_{2\nu}(\Delta E,Z) \neq G_{0\nu}(\Delta E,Z)$ and the matrix elements
$M^{F}_{2\nu}$ and $M^{GT}_{2\nu}$, which don't contain the neutrino potential
$H$ but do contain energy denominators because, without the intermediate
neutrino, closure is not a good approximation.  For later reference,
we give the
relevant \BBt\ matrix elements here:
\begin{eqnarray}
   M^{F}_{2\nu}&=&\sum_m \frac{\langle f |\sum_j \tau^+_j |m\rangle \langle m| 
\sum_k
   \tau^+_k |i\rangle}{E_m-[M_i+M_f]/2}\\
   M^{GT}_{2\nu}&=&\sum_m \frac{\langle f |\sum_j \vec{\sigma}_j \tau^+_j
|m\rangle \langle m|\sum_k
   \vec{\sigma}_k \tau^+_k |i\rangle}{E_m-[M_i+M_f]/2} ~.
\end{eqnarray}

Other mechanisms besides light-neutrino exchange can drive \BBz\
decay, and we discuss them briefly later.  No matter what the
exchanged particles, however, the occurrence of \BBz\ decay implies
that neutrinos
are Majorana particles with nonzero mass (Schechter and Valle  1982).

\section{Phenomenology of Neutrino Properties and Double-Beta Decay}
%
%

The remarkably successful worldwide neutrino physics program has
revealed much about neutrinos
over the past decade. We now know that they mix and we have
initial values for the mixing matrix elements.
We know the differences between the squares of the neutrino masses
and the number of light active neutrino
species. There is much we still don't know, however. In this section
we summarize what we have learned and its implications for
\BB\-decay experiments that seek to learn more, focusing as 
mentioned above on the  exchange of three light species of neutrinos. 
Other \BB\ possibilities are discussed later.

What aspects of
still-unknown neutrino
physics is it  most important to explore?
Although the answer is to a certain degree a matter of opinion,
it is clear that the absolute mass scale and whether the
neutrino
is a Majorana or Dirac particle
are crucial issues. \BB\ decay is the only laboratory process that can
test the absolute mass scale with a
sensitivity near $\sqrt{\delta m_{\rm atm}^{2}}$ (defined and discussed
below). More importantly, whether the neutrino is Majorana
or Dirac is a completely open question, and \BB\ decay is the only practical
way
to address it. Because future \BB-decay
experiments will be sensitive to
a range of masses that includes $\sqrt{\delta m_{\rm atm}^{2}}$ and
therefore at least one neutrino, even null
results will have significant impact on our understanding. We believe
that \BBz\ decay should be part of any future experimental neutrino program.

\subsection{Neutrino-Oscillation Parameters}
   %
   %

In this subsection, we define the various neutrino-oscillation parameters and
discuss the ways
their uncertainties affect our ability to extract \mee\ from a \BB\ 
experiment.
The oscillation experiments have provided data on the mixing-matrix elements
and the differences in the squares of the mass eigenvalues ($\delta
m_{ij}^2 \equiv ( m_j^2 - m_i^2)$). From the atmospheric-neutrino
data, we have $|\delta m_{23}^{2}| \equiv \delta m_{\rm atm}^{2} =
2.0_{-0.7}^{+1.0}
\times 10^{-3}$ eV$^2$(90\% CL) and $\theta_{\rm atm} \equiv \theta_{23}
\approx 45$
degrees (Saji 2004). The combined results of solar-neutrino
experiments and reactor experiments (Ahmed \etal  2004) give
$\delta m_{12}^{2} \equiv \delta m_{\rm sol}^{2} = 7.1_{-0.6}^{+1.2}
\times 10^{-5}$ eV$^2$  (much less than $\delta m_{\rm atm}^{2}$) and
$\theta_{12} \equiv \theta_{\rm sol} = 32.5_{-2.3}^{+2.4}$ degrees
(68\% CL).  From reactor experiments, we have the limit
$\theta_{13} < 9$ degrees (68\% CL) (Hagiwara 2002).
Note that other authors obtain modestly different
results (see, {\it e.g.}
Bahcall and Pe\~{n}a-Garay 2003, de Holanda and Smirnov 2003, 
Smy \etal  2004)
for the solar-reactor parameters. The Super-Kamiokande collaboration has
also presented an independent value for $|\delta m_{23}^{2}|$ 
of $2.4 \times 10^{-3}$ 
eV$^2$ (Ishitsuka 2004).
Finally the limits on $\theta_{13}$ depend on the specific value of 
$\delta m_{23}^{2}$.
For these reasons and the fact that the precise values for these parameters are
rapidly changing, the discussion to follow should be considered illustrative 
only.

The sign of  $\delta m_{\rm sol}^{2}$ is known; the lighter of the
two eigenstates participating significantly in the solar
oscillations, which we call $\nu_1$,
has the largest $\nu_e$ component. (The third eigenstate $\nu_3$ contains
very little $\nu_e$.) We know that $\nu_1$ is slightly lighter than
$\nu_2$, but we don't know whether $\nu_3$ is heavier or lighter than
this pair.  If $\nu_3$ is heavier, the arrangement of masses is called
the ``normal hierarchy'' (with two light neutrinos and a third
significantly heavier one); if it's lighter the arrangement is called the
``inverted hierarchy''.  When all three masses are significantly
larger than $\sqrt{\delta m_{\rm atm}^2}$ the hierarchy is referred to
as ``quasidegenerate'', no matter which eigenstate is the
lightest.  One of the largest questions left in the neutrino world is
``Which of the hierarchies is realized in nature?''.  The answer
will tell us about the overall scale of the neutrino masses as
well as the order in which they are arranged.

\begin{figure}
\vspace{9pt}
\begin{center}
\epsfbox{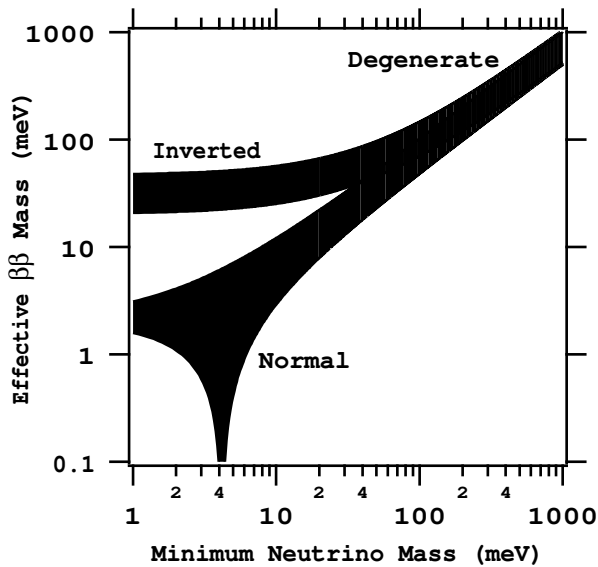}
\end{center}
\caption{The effective Majorana mass \mee\ as a function of the lightest
neutrino mass.}
\label{fig:1}
\end{figure}

The central values of the oscillation parameters
and Eqn. \ref{eq:mass} determine a range of \mee\ values for a given
value of $m_1$.  Many authors have analyzed the dependence
(see, {\it e.g.}, Vissani 1999, Bilenky \etal  1999,
Klapdor-Kleingrothaus \etal  2001, Matsuda \etal  2001, Czakon \etal
2001, Elliott and Vogel 2002,
Feruglio \etal  2002, Giunti 2003, Joaquim 2003,
Pascoli and Petcov 2003, 2004, 
Sugiyama 2003, Bilenky \etal 2004, Murayama and 
Pe\~{n}a-Garay 2004)
and Fig.\ \ref{fig:1} shows
the results. The bands indicate the range of
possible values, which depend on the unknown phases in the mixing
matrix.  The borders indicate
CP-conserving values of the
phases, $e^{i (\alpha_1-\alpha_2)} = \pm 1$.

The observation of \BBz\ decay would have profound implications
regardless of
uncertainty in the deduced value of \mee. It would show that neutrinos
are massive Majorana particles and that the total
lepton number is not a conserved quantity. But it is interesting to
ask whether one can use a measurement of \mee\ to discern the correct
hierarchy. At high values of the minimum neutrino mass, the mass spectrum is
quasi-degenerate, and the bands in Figure \ref{fig:1} are not resolved.
For a minimum neutrino mass of  about 50 meV, the degenerate band 
splits into two, representing
the normal ($m_1$ lightest) and inverted ($m_3$ lightest)
hierarchies. Figure
\ref{fig:1} appears to imply that it would be straight-forward to
identify the appropriate band at
these low mass values. However, there are uncertainties
in the oscillation parameters and the matrix elements that are not
represented in the figure.

One can address the question of how difficult it would be to distinguish
the hierarchies by comparing the maximum value for the normal
hierarchy ($\langle m_{\beta\beta} \rangle^{\rm Nor}_{\rm max}$) with
the
minimum value
for the inverted hierarchy ($\langle m_{\beta\beta} \rangle^{\rm
Inv}_{\rm min}$)
that can result from the parameter uncertainties. When the
lightest neutrino mass is small, the expressions for 
these values are simple
(Pascoli and Petcov 2003). When $m_1$
is near zero, $\langle m_{\beta\beta} \rangle^{\rm Nor}_{\rm max}$ occurs for
constructive
interference between the contributions from the
$m_2$ and $m_3$ terms in Eq.\ \ref{eq:mass}, when CP is conserved:


\begin{equation}
\label{eqn:3}
\langle m_{\beta\beta} \rangle^{\rm Nor}_{\rm max}  \approx  \sqrt{\delta
m_{\rm sol}^{2}}
\sin^2\theta_{\rm sol} \cos^2\theta_{13} + \sqrt{\delta m_{\rm atm}^{2}}
\sin^2\theta_{13}  ~.
\end{equation}
From Eq.\ \ref{eqn:3}, it is clear that $\langle m_{\beta\beta}
\rangle^{\rm Nor}_{\rm max}$ is maximal
when $\theta_{13}$, $\theta_{\rm sol}$, and the $\delta m^2$'s are as
large as they can be.

$\langle m_{\beta\beta} \rangle^{\rm Inv}_{\rm min}$ is minimal with the
same conditions on $\theta_{13}$ and $\theta_{\rm sol}$, but the smallest
allowed value for $\delta m_{\rm atm}^{2}$:

\begin{equation}
\label{eqn:4}
\langle m_{\beta\beta} \rangle^{\rm Inv}_{\rm min} = \sqrt{\delta
m_{\rm atm}^{2}}
\cos2\theta_{\rm sol} \cos^2\theta_{13}~.
\end{equation}
If we use the appropriate extremum values for the oscillation
parameters in Eqns.\ \ref{eqn:3} and \ref{eqn:4}, we find
$\langle m_{\beta\beta} \rangle^{\rm Nor}_{\rm max} \approx (9.1$
meV$)(0.327)(0.976)+ (55$ meV$)(0.024) = 4$ meV and
$\langle m_{\beta\beta} \rangle^{\rm Inv}_{\rm min} \approx (36$
meV$)(0.345)(0.976) = 12$ meV.
These numbers are sufficiently different, at least when using
our low-confidence-level uncertainty ranges, that one could
discriminate between the two
solutions.

Since the precision of the oscillation parameters is likely to improve
with future experiments, they would
not appear to be a primary concern.
Even so, it's interesting to see the effects of uncertainties in
individual parameters.  These effects can be
determined by propagating the uncertainty in each parameter through to the
uncertainty
in \mee.
The results are shown in
Table \ref{table:3}.  It is clear that $\theta_{13}$ affects
$\langle m_{\beta\beta} \rangle^{\rm Nor}_{\rm max}$ a great deal.  It's
also clear that $\delta m_{\rm atm}^{2}$ is critical for $\langle
m_{\beta\beta} \rangle^{\rm Inv}_{\rm min}$.
Finally, $\theta_{\rm sol}$  is  important for both. In short, improved
precision
for
$\theta_{13}$, $\theta_{\rm sol}$ and  $\delta m_{\rm atm}^{2}$ would help
with the interpretation of a \BBz\ experiment.

The parameter  $\theta_{\rm sol}$ has an effect in Figure \ref{fig:1} that
we haven't yet mentioned.  If it has the right value, cancellation
can drive \mee\ to very small values. But as long as
solar mixing is substantially different
from maximal, the cancellation is possible only over a narrow range of
values for the lightest mass, and complete cancellation is
not possible at all in the inverted hierarchy.

\begin{table*}
\caption{A summary of the impact on  \mee\ in the normal
and inverted hierarchies
of the oscillation-parameter uncertainties. For the central values of
the parameters, $\langle m_{\beta\beta} \rangle^{\rm Nor}_{\rm max}$
and $\langle m_{\beta\beta} \rangle^{\rm Inv}_{\rm min}$ are  2.4 meV and 19
meV, respectively. See
Pascoli and Petcov (2003) for a similar analysis.}
\label{table:3}
\newcommand{\m}{\hphantom{$-$}}
\newcommand{\cc}[1]{\multicolumn{1}{c}{#1}}
\renewcommand{\tabcolsep}{2pc} 
\renewcommand{\arraystretch}{1.2} 
\begin{tabular}{@{}cccc@{}}
\hline
Oscillation                  & Parameter        & Range 
&     Range                             \\
Parameter                    &   Range          & in  $\langle
m_{\beta\beta} \rangle^{\rm Nor}_{\rm max}$ & in $\langle
m_{\beta\beta}
\rangle^{\rm Inv}_{\rm min}$       \\
\hline
$\sqrt{\delta m_{\rm sol}^{2}}$  & 8.1 - 9.1 meV    &      2.3 - 2.6 
meV                       &       N.A. 
\\
$\sqrt{\delta m_{\rm atm}^{2}}$  & 36 - 55 meV      &   3.2 - 3.7 meV 
&       15.2 - 23.2 meV                               \\
                              &                  &   (with 
$\theta_{13}=9^o$)               & 
\\
$\theta_{\rm sol}$               & 30.1 - 34.9 deg  &   2.1 - 2.7 meV 
&       15.5 - 22.4 meV                          \\
$\theta_{13}$                & 0 - 9 deg        &   2.4 - 3.5 meV 
&       18.6 - 19.0 meV                          \\
\hline
\end{tabular}\\[2pt]
\end{table*}

The uncertainty in \Mz\ has been a source of concern for a long time. 
Typically, it has
been assumed to contribute of a factor of 2--3 times \mee\
to the uncertainty in \mee. This
uncertainty clearly
dwarfs any from the oscillation parameters and thus is the primary
issue. We address it later.

\subsection{The Absolute Mass Scale }
As already mentioned, we know that at least one neutrino has a mass
greater than $\sqrt{\delta m_{\rm atm}^{2}} \sim 45$ meV; sensitivity 
to  this value is therefore
the goal of most future neutrino-mass experiments.  In this 
subsection we  briefly compare the potential
of other mass measurements.

There are many ways to measure the mass of the neutrino.
(see Bilenky \etal  2003) for a nice summary
of the techniques). The best of these are \BBz decay, $\beta$ decay, and 
cosmological observations. These three approaches
are complementary in that they determine different combinations of 
the mass eigenvalues and mixing-matrix parameters
(see Equations \ref{eqn:massformulas}). A measurement of \BBz\ decay 
determines a coherent sum of the Majorana neutrino masses because 
\mee\
arises from exchange of a virtual neutrino. Beta decay measures an 
incoherent sum because a real neutrino is emitted.
The cosmology experiments measure the density of neutrinos and thus a 
parameter proportional to the sum of the neutrino masses.

The present limit \mb\ $\leq 2200\ $meV $ (95\% CL)$ comes from 
tritium beta decay (Lobashev \etal  1999, and Weinheimer \etal  1999).
This limit, when combined with the oscillation results, indicates 
that for at least one neutrino:
\be
   45\ {\rm meV} \leq m_i \leq 2200\ {\rm meV}~.
\ee
The two $\beta$-decay experimental groups just referenced have joined 
forces to form the KATRIN collaboration (Osipowicz \etal  2001).
They propose to build a large spectrometer that exploits the 
strengths of both previous efforts.  KATRIN hopes to reach a 
sensitivity
to \mb\ near 200 meV (Bornschein 2003).

Massive neutrinos would contribute to the cosmological matter density 
(Hannestad 2003) an amount,
\be
   \Omega_{\nu}h^2 = \Sigma/92.5\ {\rm eV}~,
\ee
where $\Omega$ is the neutrino mass density relative to the critical 
density, $100 h$ is the Hubble constant in km/s/Mpc, and $\Sigma 
\equiv m_1 + m_2 + m_3$ is the sum of the neutrino masses. The 
neutrinos are light,
however, and cluster with cold dark components of the matter density 
only for scales larger than
\be
   k \sim 0.03 \sqrt{ (m_{\nu}/1 {\rm eV}) \Omega_m} h\ {\rm Mpc}^{-1}.
\ee
For smaller values of $k$, perturbations are suppressed, and so 
measurements of the large
scale structure (LSS) can provide constraints on the neutrino mass. 
Such constraints are
rather weak, though, unless used in conjunction with precise 
determinations of the other
various cosmological parameters, which also reflect the size of 
perturbations. Most
recently the WMAP collaboration (Bennett \etal  2003) has provided 
precise cosmological
data that supplement LSS data from the 2dF galaxy survey (Elgar{\o}y 
\etal  2002), CBI
(Person \etal  2002), ACBAR (Kuo 2004) and the Lyman-$\alpha$ forest 
data (Croft 2002).
These data have been used in various combinations to derive 95\% CL 
limits on $\Sigma$
of $< 0.75$ eV (Barger \etal  2003), $< 1.0$ eV (Hannestad 2003a), $< 
1.7$ eV (Tegmark \etal  2003),
  $< 0.69$ eV (Spergel 2003), and $<1.0$ eV (Crotty \etal  2004). 
There is at least one claim for a nonzero value: $\Sigma = 0.64$ eV
(Allen 2003). An interesting paper by Elgar{\o}y and Lahav (2003) 
points out the impact of the prior
distributions on the resulting neutrino mass limit.

Future measurements by the Sloan Digital Sky survey (SDSS 2003) and 
the PLANCK satellite (PLANCK 2003)
may obtain limits on $\Sigma$ as low as $40$ meV (Hu 1999).  But
the determination of $\Sigma$ from cosmology is clearly complicated by
the large number of correlated parameters that must be measured. 
Clean
laboratory measurements of the neutrino mass will always be desirable.
Ordinary $\beta$-decay will be hard pressed to reach a sensitivity to
$\sqrt{\delta m_{\rm atm}^{2}}$, so \BBz\ experiments are especially important.

\subsection{The Majorana Phases}
\begin{figure}
\vspace{9pt}
\begin{center}
\epsfbox{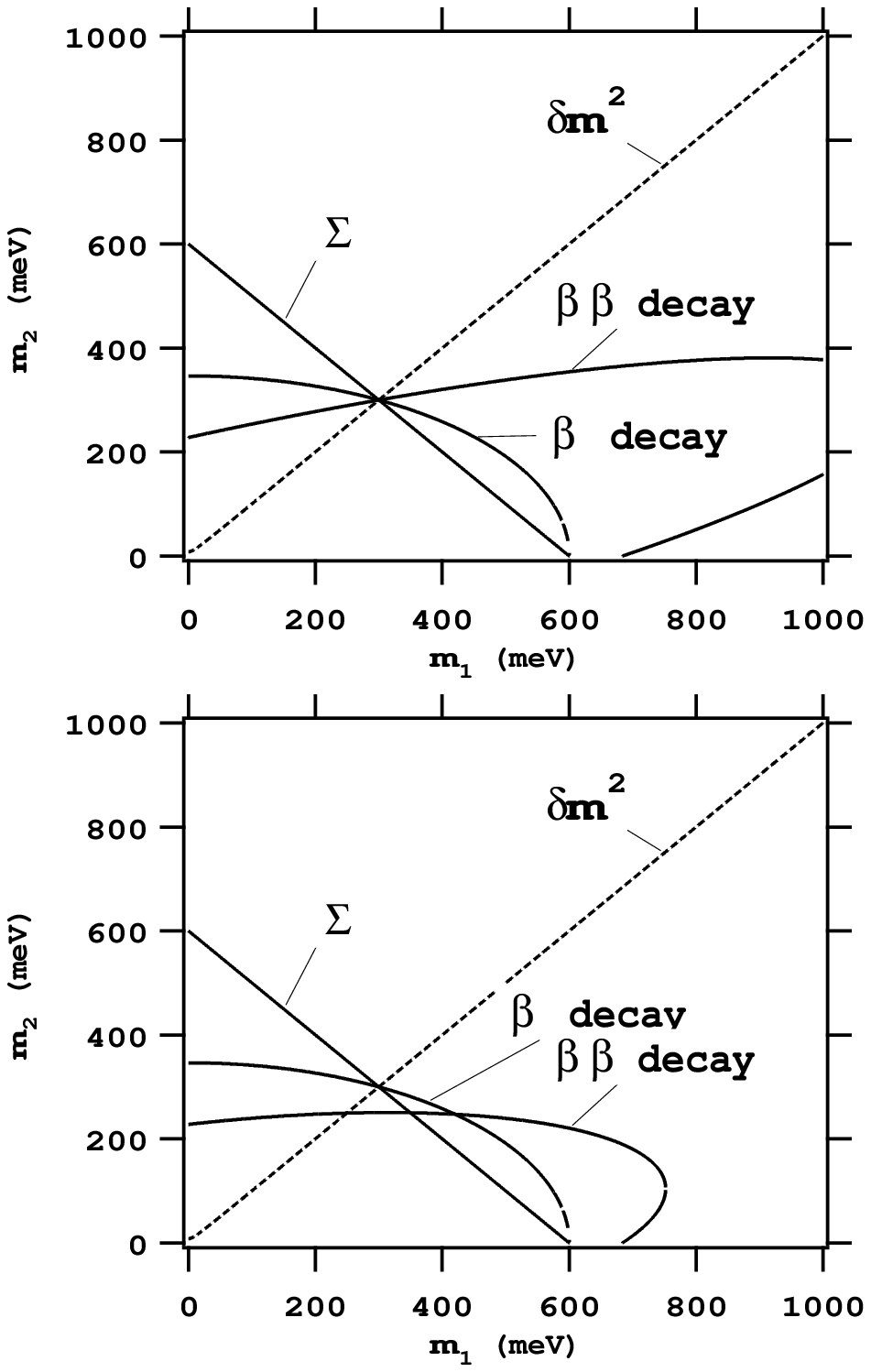}
\end{center}
\caption{A consistency plot for the neutrino mass eigenvalues $m_1$ 
and $m_2$, for various
hypothetical measurements. This set of curves indicates how measured 
values of $\Sigma$,
\mee, \ms, and \mb\ constrain the mass eigenvalues. The \BBz\ curve 
has been drawn for an {\em incorrect}
value of the phase in the bottom panel to indicate the sensitivity of 
this technique for extracting the CP-violating phase.
See text for further description of the parameters
used to draw the curves.}
\label{fig:masses}
\end{figure}

Equation \ref{eq:mass} shows the effective Majorana neutrino mass and 
its relation to the Majorana phases.
When these relative phases are an integer multiple of $\pi$ CP is 
conserved.  In principle, the two relative
phases ($\alpha \equiv (\alpha_2 - \alpha_1)$, $\beta \equiv 
(-\alpha_1 - 2\delta)$) have measurable
consequences. In practice, determining them will be difficult. In 
this subsection, we discuss the
physics of these phases.
Many authors have examined the potential to combine measurements from 
\BB\ decay, tritium $\beta$
decay, and cosmology to determine the Majorana phases. (See for 
example: Sugiyama 2003; Abada and Bhattacharyya 2003;
Pascoli \etal 2002, Pascoli \etal 2002a.) We can illustrate the main 
ideas through a 
simplified set of hypothetical
measurements. Figure \ref{fig:masses} shows
a two-neutrino-species example of such a set. We took the mixing 
matrix and $\delta m^2$ to be the best fit to the
solar-neutrino data, with an arbitrary 
value for the Majorana phase $\alpha$
(of which there is only one) of 2.5 radians.  We then made up values 
for $\Sigma$, \mee, and \mb\, assuming them to be the results of pretend
measurements.  Each curve in 
the $m_2$ vs.\ $m_1$ graph is defined by
one of these measurements.  We chose the value of $\Sigma$ (from cosmology)
to be 600 meV, corresponding to a quasidegenerate hierarchy, and let 
\mb\ = 300 meV and \mee\ = 171 meV.
The $m_2$ versus $m_1$ curves from the ``measurements" of the 
oscillation parameters, $\Sigma$, and $\beta$ decay are:
\begin{eqnarray}
  m_2 & = & \Sigma - m_1 \nonumber \\
  m_2 & = & \sqrt{m_1^2 + \delta m_{12}^{2}}  \nonumber \\
  m_2 & = & \sqrt{ (\langle m_{\beta} \rangle/U_{e2})^2 - (m_1 
U_{e1}/U_{e2})^2}~.
\end{eqnarray}

The \BB\ constraint is a little more complicated than that from 
$\beta$ decay. The curve is also
an ellipse but rotated with respect to the axes.
The constraint is given by the solution to the quadratic equation
resulting from the two-neutrino version of Equation \ref{eq:mass}:
\begin{eqnarray}
a & = & U_{e2}^4  \nonumber \\
b & = & 2 m_1 U_{e1}^2 U_{e2}^2 \cos(\alpha)  \nonumber \\
c & = & (m_1 U_{e1}^2)^2 - \langle m_{ee} \rangle  \nonumber \\
m_2 & = & (-b \pm \sqrt{b^2 - 4ac})/2a
\end{eqnarray}
All of these equations express $m_2$ in terms of $m_1$ and measured 
parameters and all should
intersect at one ($m_1$,$m_2$) point. However, because the point is 
overdetermined, the \BB\ ellipse will
intersect only for a correct choice of $\alpha$.  This provides a way to
determine $\alpha$.  In Figure \ref{fig:masses} we drew the \BB\ 
ellipse for $\alpha = 2.0$
radians and for the ``true" value of 2.5 radians
to show how the intersection does indeed depend on a correct choice 
of the phase.

\begin{figure}
\vspace{9pt}
\begin{center}
\epsfbox{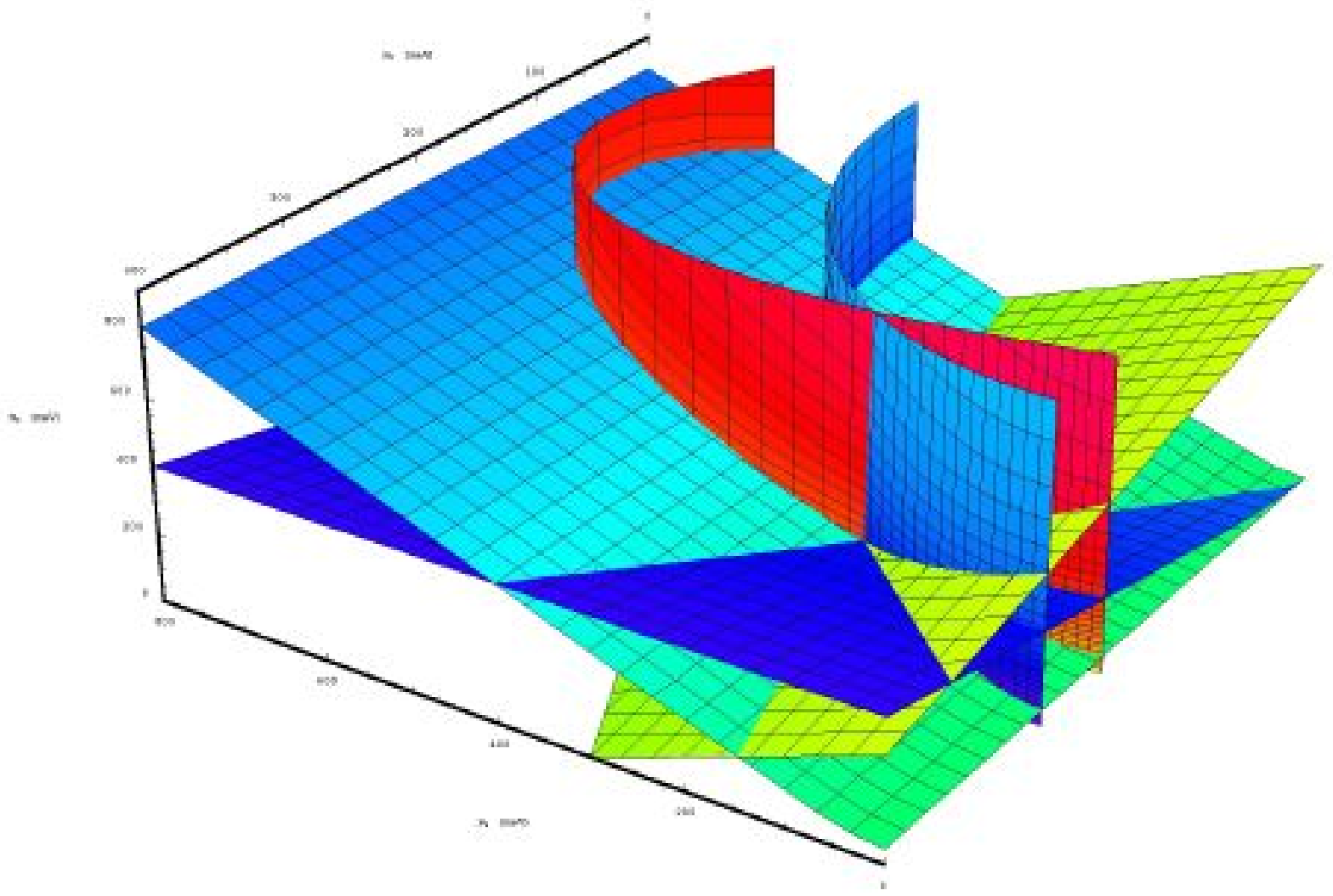}
\end{center}
\caption{A consistency plot for the neutrino mass eigenvalues $m_1$, 
$m_2$ and $m_3$ for various
hypothetical measurements. This set of surfaces indicates how measured 
values of $\Sigma$,
\mee, oscillations and \mb\ constrain the mass eigenvalues. See text for
further description of the parameters
used to draw the surfaces.}
\label{fig:masses3d}
\end{figure}

Although this two-species example is illuminating, it is overly 
simplistic. One needs to consider
three species and two phases. In this case, the $\Sigma$ constraint 
becomes a plane and the \BB- and $\beta$-decay
constraints become ellipsoids. Figure \ref{fig:masses3d} shows a 
model three-species analysis with $\Sigma = 700$ meV,
$\delta m_{32}^{2}$ positive, \mb\ = 232 meV, and \mee\ = 159 meV.
(The phases were taken to be 2.0 and 2.5 radians, and $U_{e3}$ was 
taken to be = 0.03.)
The surfaces, shown in Figure \ref{fig:masses3d}, are defined by the 
equations for $\Sigma$,
oscillations, \mb\ and \mee.  Respectively, these are:
\begin{eqnarray}
\label{eqn:massformulas}
  \Sigma & = & m_1 + m_2 + m_3  \nonumber \\
  \delta m_{21}^{2}  & = & m_2^2 - m_1^2   \nonumber \\
  \delta m_{32}^{2}  & = & m_3^2 - m_2^2   \nonumber \\
  \langle m_{\beta} \rangle^2 & = & m_1^2 U_{e1}^2 + m_2^2 U_{e2}^2 + 
m_3^2 U_{e3}^2  \nonumber \\
  \langle m_{ee} \rangle^2 & = & m_1^2 U_{e1}^4 + m_2^2 U_{e2}^4 + m_3^2
  U_{e3}^4 + 2 m_1 m_2 U_{e1}^2 U_{e2}^2 \cos(\alpha)  \nonumber \\
                           & + & 2 m_1 m_3 U_{e1}^2 U_{e3}^2 \cos(\beta) +
                           2 m_2 m_3 U_{e2}^2 U_{e3}^2 \cos(\alpha+\beta)~.
\end{eqnarray}
These surfaces intersect at a point but \BB\ decay is the only 
measurement of those used that
is sensitive to the phases. Thus a second pair of phases will
also produce a consistent result. (That additional ellipsoid is not 
shown here.)  Two experiments that depend on
the phases are required to unambiguously determine both. Furthermore, 
to keep the plots legible, this analysis ignores
any uncertainty in the measured parameters.  The uncertainties will 
be large, at least initially, and therefore
conclusions about the phases will be
weakened. Barger \etal (2002), for example, have noted that a significant
improvement in \Mz\ is necessary
before the interpretation can be successful. The articles
listed above consider the uncertainties in their analyses.

When heavy right-handed Majorana neutrinos decay, they will violate 
lepton number. In the early universe,
these decays would be out of equilibrium and could violate CP. The 
resulting net lepton number could be transferred to a
net baryon number through standard weak interactions.
Thus, this leptogenesis process (Fukugita and Yanagida 1986) could 
explain the baryon asymmetry in the Universe. In
principle, leptogenesis depends on the Majorana phases, so 
its understanding might provide the needed additional
constraint. The baryon asymmetry can be expressed (Buchm\"{u}ller 
\etal  2002) in terms of the mass $M_1$
of the lightest of the heavy Majorana neutrinos $\Phi_1$, the CP 
asymmetry $\epsilon$ in $\Phi_1$ decays,
an effective neutrino mass $\tilde{m}$, and the
sum of the squares of the three light neutrino masses. Unfortunately, the
relationship between $\epsilon$ and the low-energy phases
relevant for \BB\ is model dependent.
Many models have been studied in the literature, each with its own
relationship and conclusions.

\section{Calculating Nuclear Matrix Elements}
The observation of \BBz\ decay would immediately tell us that neutrinos are
Majorana particles and give us an estimate of their overall mass 
scale but without
accurate calculations of the nuclear matrix
elements that determine the decay rate it will be difficult to reach 
quantitative
conclusions about masses and hierarchies.  Theorists have tried hard to develop
many-body techniques that will allow such calculations.  They have tried to
calibrate their calculations to related observables:  \BBt\ decay, ordinary
$\beta^+$ and $\beta^-$ decay, Gamow-Teller strength distributions, odd-even
mass differences, single-particle spectra.  They have
tried to exploit approximate isospin and $SU(4)$ symmetries in the nuclear
Hamiltonian, to extend well-known many-body methods in novel ways.   In spite
of all this effort, we know the matrix elements with limited accuracy.
In this section we review the state of the nuclear-structure
calculations and discuss ways to improve them.   While increased
accuracy is not easy to achieve, we will argue that it is also not impossible.

Most recent attempts to calculate the nuclear matrix elements have been based
on the neutron-proton Quasiparticle Random Phase Approximation (QRPA) or
extensions to it. Of those that haven't, the most prominent are based on the
shell model. While the two methods have much in common --- their
starting point is a Slater determinant of independent particles ---
the kinds of correlations they include are complementary.  The QRPA treats a
large fraction of the nucleons as ``active" and allows these nucleons a large
single-particle space to move in.  But RPA correlations are of a specific and
simple type best suited for collective motion.  The shell model, by contrast,
treats a small fraction of the nucleons in a limited single-particle space, but
allows the nucleons there to correlate in arbitrary ways.  That these very
different approaches
yield similar results indicates that both capture most of the important
physics.

\subsection{QRPA}

The QRPA was developed by Halbleib and Sorenson (1967)
and
first applied to double-beta
decay by Huffman (1970).  Both it and early shell model
calculations had
problems reproducing measured \BBt\ rates until the realization by
Vogel and Zirnbauer (1986) that in the QRPA the neutron-proton ($np$)
particle-particle (i.e.\
pairing) interaction, which has little effect on the collective Gamow-Teller
resonance, suppresses \BBt\ rates considerably.  Soon afterward,
Engel \etal (1988) and Tomoda and Faessler (1987) demonstrated a
similar though smaller effect on \BBz\ decay.  It was
quickly realized, however, that the QRPA was not designed to handle realistic
$np$ pairing; the calculated half-lives were unnaturally sensitive to the
strength of the pairing interaction.  As a result, the rates of \BB\ decay,
particularly \BBt\ decay, were hard to predict precisely because a
small change in a phenomenological parameter (the strength of $np$
pairing)
caused a large
change in the lifetimes and eventually the breakdown (called a ``collapse") of
the entire method when the parameter exceeds some critical value.  Most recent
work in the QRPA has aimed at modifying the undesirable aspects of the
method so that its sensitivity to $np$ pairing becomes more realistic.

One approach is ``Second" QRPA (Raduta \etal  1999) . The QRPA itself 
treats states in the
intermediate nucleus as one-quasineutron + one quasiproton 
excitations of the initial and/or final ground states.  The 
quasineutron-quasiproton pair is taken to obey boson statistics, 
inspiring the term ``quasiboson
approximation" for the method.  Second QRPA extends the structure of 
these states by adding another term in a boson expansion of the pair.

A larger number of papers have been based on the Renormalized QRPA (RQRPA)
(Hara 1964, Rowe 1968,  and Suhonen 1995).  The quasiboson
approximation is equivalent to replacing commutators
by their expectation values in an independent-quasiparticle 
approximation to the
initial and final ground states.  The RQRPA uses the QRPA ground states to
evaluate the commutators.  Because the commutators in turn help fix the ground
states, the two must be evaluated self-consistently, usually via iteration.  A
variant of this approach is the ``Full RQRPA", in which the effects 
of isovector
$np$ pairing, artificially strengthened to account implicitly for isoscalar
pairing, are included in the BCS calculation that defines the quasiparticles as
well as in the subsequent QRPA calculation (Schwieger \etal  1996; \v{S}imkovic
\etal  1997).  [Isovector $np$ pairing
was first introduced in this way in the unrenormalized QRPA (Cheoun \etal 
1993, 1995, Pantis \etal 1996)].
Another extension is the Self-Consistent RQRPA (SCQRPA).  Here, the BCS
calculation is modified and iterated together with the RQRPA 
calculation until the RQRPA ground state has the same number of 
quasiparticles as the BCS-like state on which it is based.  All these 
methods reduce the dependence of the \BBt\ matrix elements on the 
strength of the neutron-proton pairing interaction.

Unfortunately, it's not clear which of the methods is best.  One might presume
that the Full RQRPA and the Self-Consistent RQRPA are better than the vanilla
RQRPA, which in turn is better than the original unrenormalized QRPA. But the
results of the RQRA and Full RQRPA appear to be quite sensitive to, e.g., the
number of single-particle states in the model space (\v{S}imkovic \etal 
1997). They also
violate an important sum rule for single $\beta$ strength, and studies in
solvable models suggest that the reduced dependence (at least of the RQRPA) on
neutron-proton pairing may be spurious (Engel \etal  1997) resulting from
an artificial
reduction of isoscalar pairing correlations.  And it is not clear that the Full
RQRPA's substitution of isovector pairing for isoscalar pairing is legitimate.

The Self-Consistent QRPA has been
applied to \BBz\ decay only once, and that application did not take into
account the predictions of the same approach for \BBt\ decay.  The
approximations that go into the Second
QRPA are of a different type, and while one study indicates less dependence
on model space size for that method (Stoica and Klapdor-Kleingrothaus 2001)
there are not many other
reasons to prefer one approach or the other.

Recently, the RQRPA was modified even further, so that the single $\beta$
sum rule was restored.
The resulting method, called the ``Fully Renormalized QRPA"
has yet to be applied to \BBz\ decay.  Even more recently, \v{S}imkovic
\etal (2003)
raised the issue of nuclear deformation, which has usually been ignored in
QRPA-like treatments of nearly spherical nuclei\footnote{Psuedo-SU(3)-based
truncations have been used to treat it in well-deformed nuclei ({\it 
e.g.}, in Hirsch {\it et al}
1996). }. The authors argued that
differences in deformation between the initial and final nuclei can 
have large effects on the \BBt\ half-life. These ideas, too, have not 
yet been
applied to \BBz\ decay.

The profusion of RPA-based acronyms is both good and bad.  The sheer 
number of methods applied gives us a kind of statistical sample of 
calculations, which we will use below to get an idea of the 
theoretical
uncertainty in the matrix elements.  But the sample may be biased by the
omission of non-RPA correlations in all but a few calculations.  Other
approaches that include correlations more comprehensively should be pursued.

\subsection{Shell Model}

The obvious alternative to RPA, and the current method of choice for nuclear
structure calculations in heavy nuclei where applicable, is the shell model.
It has ability to represent the nuclear wave function to arbitrary accuracy,
provided a large enough model space is used.  This caveat is a huge one,
however.
Current computers allow very large bases (millions of states), but in heavy
nuclei this is still not nearly enough.  Techniques for constructing
``effective'' interactions and operators that give exact results in truncated
model spaces
exist but are hard to implement.  Even in its crude form with relatively small
model spaces and bare operators, however, the shell model offers advantages
over the QRPA.  Its complicated valence-shell correlations, which the QRPA
omits (though it tries to compensate for them by renormalizating parameters)
apparently affect the \BB\ matrix elements (Caurier \etal  1996).

The first modern shell-model calculations of \BB\ decay date from reference
Haxton and Stephenson (1984) and references therein.   Only a few truly
large-scale shell model calculations have been
performed.  The heavy deformed \BB\ nuclei, $^{238}U$, and 
$^{150}$Nd, for example,
require bases that are too large to expect real accuracy.  Realistic work has
thus been restricted to $^{48}$Ca, $^{76}$Ge, and $^{136}$Xe, though less
comprehensive calculations have been carried out in several other nuclei
(Suhonen \etal  1997).

Large spaces challenge us not only through the problem of diagonalizing large
matrices, but also by requiring us to construct a good effective interaction.
The bare nucleon-nucleon interaction needs to be modified in truncated spaces
(this is an issue in the QRPA as well, though a less serious one). 
Currently, effective interactions are
built through a combination of perturbation theory, phenomenology, and
painstaking fitting.  The last of these, in particular, becomes increasingly
difficult when millions of matrix elements are required.

Related to the problem of the effective interaction is the renormalization of
transition operators.  Though the problem of the effective 
Gamow-Teller operator
(Siiskonen \etal  2001), which enters directly into \BBt\ decay, has drawn
some attention,
very little work has been done on the renormalization of the two-body operators
that govern \BBz\ decay in the closure approximation.  Shell model 
calculations won't be truly reliable until they address this issue, 
which is connected with deficiencies in the wave function caused by 
neglect of single-particle levels far from the Fermi surface.  Engel 
and Vogel (2004) suggests that significant
improvement on the state of the art will be difficult but not 
impossible in the coming years.

\subsection{Constraining Matrix Elements with Other Observables}

The more observables a calculation can reproduce, the more trustworthy it
becomes.  And if the underlying model contains some free parameters, these
observables can fix them.  The renormalization of free parameters can make up
for deficiencies in the model, reducing differences between, e.g., the QRPA
and RQRPA once the parameters of both have been fit to
relevant data.  The more closely an observable resembles \BBz\ decay, the more
relevant it is.

Gamow-Teller distributions, both in the $\beta^-$ and $\beta^+$ directions,
enter indirectly into both kinds of \BB\ decay, and are measurable through
$(p,n)$ reactions.   Aunola and Suhonen (1998) are particularly careful to
reproduce those transitions as well as possible.  Pion double charge exchange,
in which a $\pi^+$ enters and a $\pi^-$ leaves, involves the transformation of
two neutrons into two protons, like \BB\ decay, but the nuclear operators
responsible aren't the same in the two cases. Perhaps the most relevant
quantity for calibrating calculations of \BBz\ decay is \BBt\ decay, which has
now been measured in 10 different nuclei.

Two recent papers have tried to use \BBt\ decay to fix the strength of $np$
pairing in QRPA-based calculations.  Stoica and Klapdor-Kleingrothaus
(2001) used it only for
the
$J^{\pi}=1^+$ channel relevant for \BBt\ decay, leaving the $np$ pairing
strength unrenormalized in other channels.  By contrast, Rodin \etal (2003)
renormalized the strength in all channels by the same amount.  The results of
the two procedures were dramatically different:  Stoica and
Klapdor-Keingrothaus (2001)
found that the \BBz\ matrix elements depended significantly on the theoretical
approach
(QRPA, RQRPA, FRQRPA, second QRPA) and, in some of the approaches, on the model
space, while Rodin \etal (2003) found almost no dependence on model-space
size, on the form of the nucleon-nucleon interaction, or on whether the QRPA or
RQRPA was used.  The authors argued that fixing the $np$ pairing strength to
\BBt\ rates essentially eliminates uncertainty associated with variations in
QRPA calculations of \BBz\ rates, though they left open the question of how
close to reality the calculated rates were.

Given all these considerations, can we meaningfully estimate the uncertainty
in the \BBz\ matrix elements?  We address this question for a particular
nucleus now.

\subsection{How Well Can We estimate Uncertainty? The Case of $^{76}$Ge}

How accurate are existing calculations?  One can answer
provisionally by looking at the spread in the many predictions
offered recently in the literature.  We focus here on $^{76}$Ge, which
currently gives the strongest limit on \mee\ (or perhaps even a value 
for it: see
discussions below),
and figures prominently in several proposals for new experiments.

\Tref{tab:pred}, taken in part from Civitarese and Suhonen (2003),
shows predictions
of most of the calculations in the literature for the {\it nuclear}
contribution to
the decay rate (Doi \etal  1985), $C_{mm} \equiv \left[\langle m_{\beta\beta}
\rangle ^2
T^{0\nu}_{1/2} /m_e^2\right]^{-1} = G_{0\nu}|M_{0\nu}|^2 m_e^2$, and for the
effective neutrino mass \mee\ that would be deduced by
measuring a lifetime $T^{0\nu}_{1/2}=4.0 \times 10^{27}$ years.  Besides
differing in method (by which they are grouped) the calculations use different
model spaces, fit different observables, and adjust different parameters.  The
resulting $C_{mm}$'s vary by nearly two orders of magnitude, leading to an
order of magnitude variation in the extracted neutrino mass. But some of the
``outliers'', labeled with asterisks in the table, can be eliminated on the
grounds that the corresponding
calculations omit or misrepresent clearly important effects. The VAMPIR model
used in the early calculation of Tomoda \etal (1986)
contains no $np$
pairing correlations and so gives too large a decay rate.  The shell-model
truncation by Haxton and Stephenson (1984) was done in a way that
minimized $np$
pairing, and the upper limit in $C_{mm}$ from Engel \etal  (1989)
was considered probably too large by the authors (though in their calculation
they set $g_A=1$, which here would move the upper limit into
the middle of the range).
At the other end of the
spectrum, the very small self-consistent RQRPA decay rates of reference
Bobyk \etal (2001) were obtained with a value of the $np$ pairing strength
that was not consistent with the measured \BBt\ rate; when the strength
is adjusted to reproduce \BBt\ decay, the results for the \BBz\ rate are
close to those of
the plain QRPA in the same reference.  Without any further culling, the
remaining $C$'s vary by about 1 order of magnitude, and the extracted
\mee's vary by a factor of about three, from 0.022 to 0.068 eV for the
lifetime we have chosen.

Even if some of the other calculations are objectionable, it is difficult to
reduce the spread much below this factor of three without some real work.
Aunola and Suhonen (1998), who among all the entries do the most extensive
job of
adjusting parameters to reproduce spectroscopic data and single-$\beta$
transition strengths, obtain a low neutrino mass: 0.022 eV from our 
hypothetical
lifetime.  Caurier \etal  (1996), who do the most complete shell-model
calculation,
obtain 0.059 eV, close to the maximum mass from calculations we
haven't removed.  We cannot discount these two careful calculations without
examining them more closely
and so
cannot further restrict the range.  This is not to say that these calculations
can't be questioned.  It may not be appropriate, for example, to fit
single-particle energies to $\beta$-decay rates as Aunola and Suhnen do.  But
for now, until someone
investigates the situation further,
the range of reasonable results
should include
these two calculations.

\begin{table}
\caption{\label{tab:pred}Predictions for $^{76}$Ge.  The quantity $C_{mm}$,
defined in the text, is a measure of the nuclear contribution to the 
\BBz\ decay
rate.  The values of \mee\ are what those that would be extracted by a
calculation if the lifetime were $4.0 \times 1-^{27}$ years.
The asterisks indicate
approaches that omit important physics or have been superceded.}
\begin{tabular}{@{}llll}
\\
   $C_{mm} (Y^{-1})$&  \mee\  (eV)  & Method  & Reference  \\
   \\
   1.12$\times 10^{-13}$&0.024&QRPA&Muto \etal (1989), Staudt \etal (1990) \\
   6.97$\times 10^{-14}$&0.031&QRPA&Suhonen \etal (1992)\\
   7.51$\times 10^{-14}$&0.029&number-projected QRRA&Suhonen \etal (1992)\\
   7.33$\times 10^{-14}$&0.030&QRPA&Pantis \etal (1996)\\
   1.18$\times 10^{-13}$&0.024&QRRA&Tomoda (1991)\\
   1.33$\times 10^{-13}$&0.022&QRPA&Aunola and Suhonen (1998)\\
   8.27$\times 10^{-14}$&0.028&QRRA&Barbero \etal (1999)\\
   1.85-12.5$\times 10^{-14}$&0.059-0.023&QRPA&Stoica and 
Klapdor-Kleingrogthaus
  (2001)\\
   1.8-2.2$\times 10^{-14}$&0.060-0.054&QRRA&Bobyk \etal (2001)\\
   8.36$\times 10^{-14}$&0.028&QRPA&Civitarese and Suhonen (2003)\\
   1.42$\times 10^{-14}$&0.068&QRRA with $np$ pairing&Pantis \etal (1996)\\
   4.53$\times 10^{-14}$&0.038&QRPA with forbidden&Rodin \etal (2003)\\
   8.29$\times 10^{-14}$&0.028&RQRPA&Faessler and Simkovic (1998)\\
   1.03$\times 10^{-13}$&0.025&RQRRA&Simkovic \etal (1999)\\
   6.19$\times 10^{-14}$&0.032&RQRRA with forbidden&Simkovic \etal (1999)\\
   5.5-6.3$\times 10^{-14}$&0.034-0.032&RQRRA&Bobyk \etal (2001)\\
   2.21-8.83$\times 10^{-14}$&0.054-0.027&RQRPA&Stoica and 
Klapdor-Kleingrothaus
  (2001)\\
   3.63$\times 10^{-14}$&0.042&RQRPA with forbidden&Rodin \etal (2003)\\
   2.75$\times 10^{-14}$&0.049&Full RQRPA&Simkovic \etal (1997)\\
   3.36-8.54$\times 10^{-14}$&0.042-0.028&Full RQRPA&Stoica and
Klapdor-Kleingrothaus (2001)\\
   6.50-9.21$\times 10^{-14}$&0.032-0.027&Second QRPA&Stoica and
Klapdor-Kleingrothaus
  (2001)\\
   2.7-3.2$\times 10^{-15}$&0.155-143&Self-consistent QRPA$^*$&Bobyk 
\etal (2001)\\
   2.88$\times 10^{-13}$&0.015&VAMPIR$^*$&Tomoda \etal (1986)\\
   1.58$\times 10^{-13}$&0.020&Shell-model truncation$^*$&Haxton and Stephenson
(1984)\\
   6.87-15.7$\times 10^{-14}$&0.031-0.020&
   Shell-model truncation$^*$&Engel \etal (1989)\\
   1.90$\times 10^{-14}$&0.059&Large-scale shell model&Caurier \etal (1996)

\end{tabular}
\end{table}

As mentioned above, Rodin \etal (2003) argue that the variation
in QRPA
and RQRPA rates can be nearly eliminated by renormalizing a few parameters to
reproduce pairing gaps and \BBt\ decay rates.  Indeed, the table shows that the
QRPA and RQRPA numbers from that reference are close.  But it has yet to be
shown conclusively that \BBt\ decay is more important than single-$\beta$ decay
or other observables, and Engel and Vogel (2004) demonstrate that
in a two-level $SO(5)$-based model (Dussel \etal 1970), at least, 
the procedure
cannot fully eliminate model-space dependence when the QRPA is used.

\subsection{Reducing the Uncertainty}

What can be done to improve the situation?  In the near term, improvements can
be made in both QRPA-based and shell-model calculations.  First, existing
calculations should be reexamined to check for consistency.  One important
issue
is the proper value of the axial-vector coupling constant
$g_A$, which is often set to 1 (versus its real value of 1.26) in calculations
of $\beta$ decay and \BBt\
decay to account for the observed quenching of low-energy 
Gamow-Teller strength. What value should one use for
\BBz\ decay, which goes through intermediate states of all multipolarity, not
just $1^+$?  Some authors use $g_A=1$, some $g_A=1.26$, and some $g_A=1$ for
the $1^+$ multipole and 1.26 for the others. (Often authors don't reveal their
prescriptions.)  The second of these choices
appears inconsistent with the treatment of \BBt\ decay.
Since the square of $g_A$ enters
the matrix element, this issue is not totally trivial.
The striking
results of Rodin \etal suggest that an inconsistent treatment is responsible
for some
of the spread in \Tref{tab:pred}.  More and better charge-exchange experiments
would help teach us whether higher-multipole strength is also quenched.

Next, the various
versions of the QRPA should be tested against exact solutions in a solvable
model that is as realistic as possible.  The most realistic used so far are the
$SO(5)$-based model first presented by Dussel \etal (1970)
and used to study the
QRPA and RQRPA for Fermi \BBt\ decay by Hirsch \etal (1997), a
two-level version of that model used by Engel and Vogel (2004) for
the QRPA in
Fermi \BBt\ {\em and} \BBz\ decay, and an $SO(8)$-based model
(Evans \etal  1981)
used to test the QRPA and RQRPA for both Fermi and Gamow-Teller \BBt\ decay
by Engel
\etal (1997).  It should be possible to extend the $SO(8)$ model to several
sets of levels and develop techniques for evaluating \BBz\ matrix elements in
it.  All these models, however, leave out spin-orbit splitting, which weakens
the collectivity
of $np$ pairing.  A generalization of pseudospin-based models like those of
Ginocchio (1980), used in the Fermion Dynamical Symmetry Model
(Wu \etal  1994),
might be useful provided techniques like those of Hecht (1993) to
evaluate expectation values of operators outside the algebra can be extended.
Such calculations should help us understand the virtues and deficiencies of
QRPA extensions.

Along the same lines, we will need to understand the extent to which such
methods can reproduce other observables, and their sensitivity to remaining
uncertainties in their parameters.  A very careful study of the first issue was
made by Aunola and Suhonen (1998) and the second has been explored
in several
papers, most thoroughly by Stoica and Klapdor-Kleingrothaus (2001).
These efforts must be
extended.  The work is painstaking, and perhaps not as much fun as concocting
still more variations of the QRPA, but it is crucial if we are to reduce
theoretical uncertainty.  Self-consistent Skyrme HFB+QRPA, applied to
single-$\beta$
decay by Engel \etal (1999) and Gamow-Teller resonances
by Bender \etal (2002), may be helpful here; it offers a more general
framework, in
principal anyway, for addressing the variability of calculated matrix elements.
Solvable models can be useful here too, because they can sometimes supply
synthetic data to which parameters can be adjusted (as in Engel and
Vogel (2004)).

The best existing shell-model calculation produces smaller matrix elements
than most QRPA calculations.  Computer speed and memory is now at the point
where the state of the shell-model art can be improved.  The calculation of the
\BB\ decay of $^{76}$Ge by Caurier \etal (1996) used the
$f_{5/2}p_{3/2}p_{1/2}g_{9/2}$ model space, allowing up to 8 
particles (out of a
maximum of 14) into the $g_{9/2}$ level.  Nowadays, with the help of the
factorization method (Papenbrock and Dean 2003; Papenbrock \etal  2003a), an
accurate approximation to full shell-model
calculations, we should be able to fully occupy the $g_{9/2}$ level, and
perhaps include the $g_{7/2}$ and $f_{7/2}$
levels (though those complicate things by introducing spurious center-of-mass
motion).
In addition, one can try through diagrammatic perturbation theory to
construct effective \BBz\ operators for the model space that are 
consistent with
the effective interaction.  Though perturbation theory has 
convergence problems,
the procedure should at least give us an idea of the uncertainty in the
final answers, perhaps also indicating whether result obtained from 
the ``bare''
operators is too large or too small.  Research into effective operators has
been revived in
recent years (Haxton and Song 2000) and we can hope to improve on 
diagrammatic perturbation
theory.   One minor source of uncertainty connected with renormalization 
(which also affects the QRPA) is short-range two-nucleon correlations,
currently treated phenomenologically,
following Miller and Spencer (1976).

Some {\it ab initio} calculations of nuclear properties are now 
possible (Carlson 1998).
Although researchers have obtained
accurate results only in nuclei with $A \le 12$, we have reason to
hope that ``nearly exact'' variational or coupled-cluster methods 
(Kowalski \etal  2003)
will be applied to medium-heavy nuclei such as $^{76}$Ge in
the next 5 years.
Accurate calculations in heavier nuclei are probably further off.

In short, much can be done and we would be well served by coordinated
attacks on these problems.  There are relatively few theorists working
in \BB\ decay, and their efforts have been fragmented.  More
collaborations, postdoctoral and Ph.D\ projects, meetings, etc.,
would make progress faster.  {\it There is reason to be hopeful that
the uncertainty will be reduced.}  The shell-model matrix element may be too
small because it does not include any particles outside the $fp$-shell.  These
particles, as shown by QRPA calculations, make the matrix element larger.  We
suspect that the results of a better shell-model calculation will be closer
than the best current one to the QRPA results and that, as noted above,
the spread in those
results can be reduced.  Finally, other nuclei may be
more amenable to a good shell-model calculation than Ge.  $^{136}$Xe has 82
neutrons (a magic number) making it a particularly good candidate.

\section{Other Possible Mechanisms for Double-Beta Decay}

Although the occurrence of \BBz\ decay implies the existence of
massive
Majorana neutrinos (Schechter and Valle 1982), their exchange need not be
the dominant
contribution to the decay rate.  Almost any physics that violates lepton
number can cause
\BBz\ decay.  A heavy Majorana neutrino can be
exchanged, or supersymmetric particles, or a leptoquark.  Right-handed
weak currents, either leptonic or hadronic, can cause the absorption
of an emitted virtual neutrino without the helicity flip that depends
on the neutrino mass.  These possibilities have been reviewed by
Faessler and \v{S}imkovic (1998) and Suhonen and Civitarese (1998).
But now we know that there are light neutrinos and
that next-generation \BBz\ experiments may well allow us to learn 
something new about them.  The other possibilities are
more speculative and instead of analyzing them in detail we confine ourselves
to the question of
whether it is possible, should \BBz\ decay be observed, to determine
which mechanism is responsible.

Light-neutrino exchange with right-handed currents is unique in this
regard because the
contraction of left and right-handed currents gives rise to a term
$q_{\rho}\gamma_{\rho}$ in the numerator of the neutrino propagator
(see Equation \ref{eq:lep})
that cancels in the contraction of two
left-handed or two right-handed currents.  The extra $q$ allows the
electron to be emitted in a $p$-wave as well as an $s$-wave and
introduces a contribution to the amplitude from nucleon recoil.  In
addition to increasing the phase space, these effects lead to a different
single-electron energy distribution and opening-angle dependence than
in \BBz\ decay driven by the neutrino mass.  An experiment
sufficiently sensitive
to the energies and paths of individual electrons could
therefore
determine whether right-handed currents were driving the decay.

The effects of right-handed currents can be
shown to be negligible
unless there exist some neutrinos with masses larger than a few MeV.
Heavy Majorana neutrinos, however, are generated naturally in a variety of
models via
the see-saw mechanism.
The heavy neutrinos complicate matters because they can mediate
\BBz\ decay themselves. It seems possible
within, {\it e.g.}, left-right symmetric models, for the exchange of these
neutrinos via right-handed currents to compete with or even dominate
the exchange of light-neutrinos.

The exchange of heavy particles involves short-range propagators
that give rise to decay rates
of the same form as in the mass-driven mode:  simple
two-s-wave-electron phase space multiplied by the square of an
amplitude\footnote{Two-nucleon correlations do not suppress the effects of
heavy particles, which can be transmitted between nucleons by pions
(Prezeau \etal 2003, Faessler \etal 1997).}.
The angular distributions and single-electrons spectra will
therefore be the same in all these processes.
The only way to
distinguish one from another is to take advantage of the different
nuclear matrix elements that enter the amplitudes (leading to
different total decay rates).  Unknown parameters such  as the
effective neutrino mass or the trilinear $R$-parity-violating
supersymmetric coupling (violation of $R$ parity naturally accompanies
Majorana neutrino-mass terms) also enter the rates, so several transitions
would have to be measured.  \v{S}imkovic and Faessler (2002)
argue that this is best accomplished by measuring
transitions to several final states in the same nucleus, but if the
matrix elements can be calculated accurately enough one could also
measure the rates to the ground states of several different nuclei.

The problems in determining the source of \BBz\ decay are mitigated by
constraints from other
experiments on many extra-Standard models.  Some of these constraints
will be much stronger once
the Large Hadron Collider comes on line.  If no signs of supersymmetry
(for example) appear there, then supersymmetry probably does not exist.  If
supersymmetric particles are
found, but neutralinos make it out of the detector without decaying,
then $R$-parity, which prevents the decay of supersymmetric
partners into familiar particles, will not be strongly violated.  The
trilinear $R$-parity-violating coupling could then
be ruled out as the source of double-beta decay.  The
presence of the LHC will make complementary experiments on \BBz\ decay
still more attractive.

Some theories posit the emission Goldstone bosons called
``Majorons'', together with the emission of electrons, in \BBz\ decay.
If Majorons were emitted in a detector the total energy carried by the
newly created electrons would vary from event to event. A long lifetime
for this decay would make it difficult to detect above a \BBt-decay background.
Double-beta decay has also been discussed as
as test of special relativity and the equivalence principle 
(Klapdor-Kleingrothaus \etal
1999). Finally, very recent attempts to unify the dark sector
and neutrino physics (Kaplan \etal 2004) posit a scalar field with
variations on the scale of millimeters that couples to neutrinos.
If such a field existed, the rate of \BBz\ decay might depend on the
density of matter in which the process occurred.

\section{Experimental Situation}

\begin{figure}
\vspace{9pt}
\begin{center}
\epsfbox{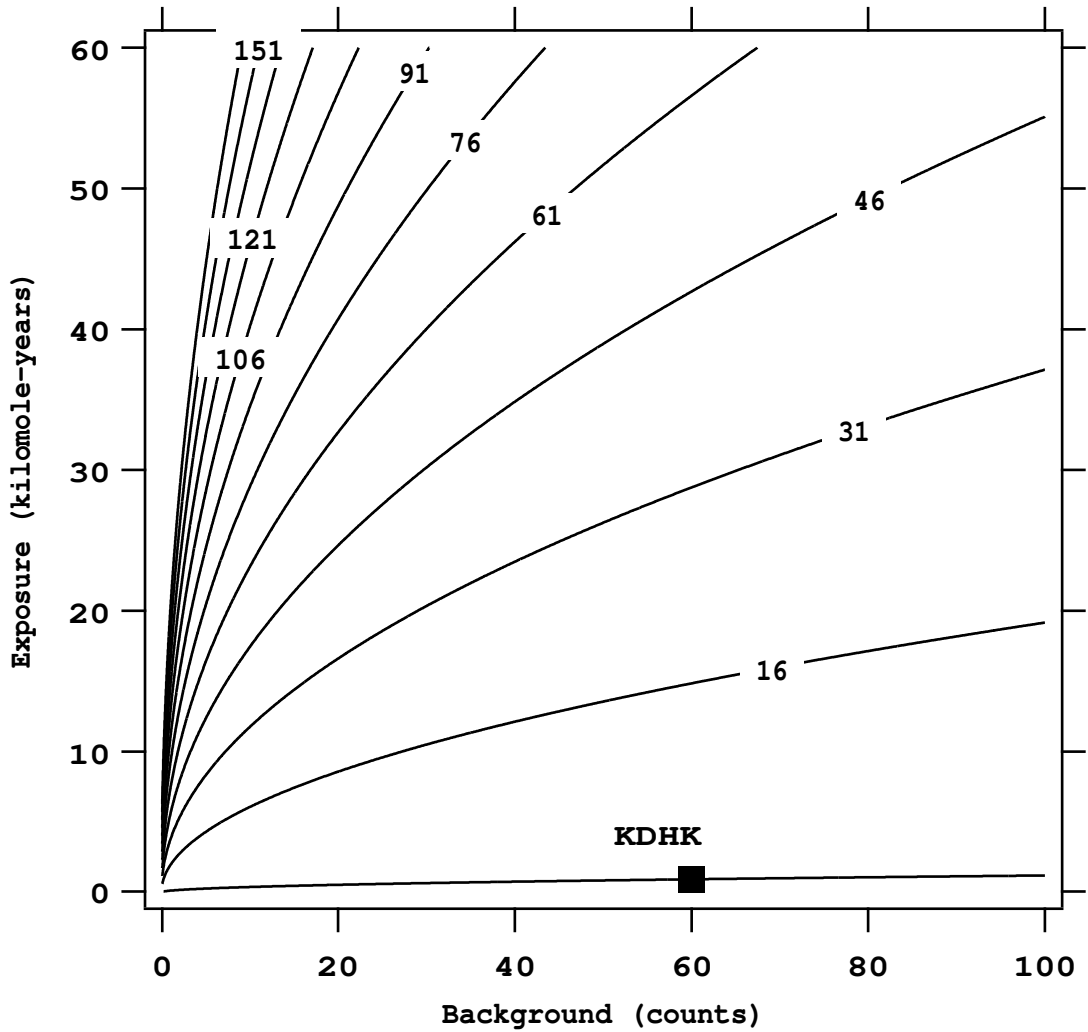}
\end{center}
\caption{This contour plot shows the half life, in units of $10^{25}$ 
y, for a peak of $5\sigma$ significance
for a given exposure and background. The KDHK point is shown.}
\label{fig:sensitivity}
\end{figure}

   If an experiment observes \BBz\ it will have profound physics 
implications. Such an
extraordinary claim will require extraordinary evidence.
The recent claim (Klapdor-Kleingrothaus \etal  2004)
for an observation of \BBz\ has been controversial (See discussion 
below). Also previous ``false peaks"
in \BB\ spectra have appeared near a \BBz\ endpoint energy (see 
discussion in Moe and Vogel 1994, page 273).
One must ask the question: What
evidence is required to convincingly demonstrate that \BBz\ has 
been observed?
Low-statistical-significance peaks ($\approx 2\sigma$)
have faded with additional data, so one must require strong 
statistical significance
(perhaps 5$\sigma$). (See Fig. \ref{fig:sensitivity}.) This will 
require a large signal-to-noise ratio that will most likely be
accomplished by an ultra-low-background experiment whose source is 
its detector. Such experiments
are usually calorimetric and provide little information beyond just 
the energy measurement.

How does an experiment demonstrate that an observed peak 
is actually due to \BB\ decay
and not some unknown radioactivity?
Additional information beyond just an energy measurement may be 
required. For example,
although there is some uncertainty
associated with the matrix elements, it is not so large that a 
comparison of measured rates in
two different isotopes could not be used to demonstrate consistency 
with the Majorana-neutrino hypothesis.
Alternatively, experiments that provide an additional handle on the 
signal, for example by measuring a variety of kinematical
variables, demonstrating that 2 electrons are present in the final 
state, observing the $\gamma$ rays associated
with an excited state, or identifying the daughter nucleus, may 
lend further credibility to a claim.
Experiments that provide this extra handle may require a 
significantly more complicated
apparatus and therefore face additional challenges.

The exciting aspect of \BB\ research today is that many proposed 
experiments intend to reach a Majorana
mass sensitivity of $\sqrt{\delta m_{\rm atm}^{2}}$. Several 
different isotopes and experimental
techniques are being pursued actively and many of the programs look viable.
In this section we describe the current situation in experimental 
\BB\ decay .

\subsection{Results to Date}
Table \ref{tab:ZeroNuResults} lists the recent \BBz\ results. The 
best limits to date come
from the enriched Ge experiments. The two experiments had comparable
results although the Heidelberg-Moscow result was marginally 
better. The \Tz\ limits near $2 \times 10^{25}$ y results in a \mee\
limit near 300 meV, with an uncertainty of about a factor of 3 
because of the uncertainty
in \Mz. One recent paper (Zdesenko \etal 2002) 
performed a joint analysis of the two experiments and
found \Tz\ $> 2.5 \times 10^{25}$ y.

Most of the results listed in Table \ref{tab:ZeroNuResults} are at least
a few years old. The obvious exceptions to this are the Te and Cd results.
CUORICINO continues to collect data.

\begin{table}
\caption{\protect  A summary of the recent \BBz\ results. The \mee\ 
limits are those
deduced by the authors. All limits are at 90\% confidence level 
unless otherwise indicated. The columns providing the
exposure and background are based on arithmetic done by the authors 
of this paper, who take responsibility for any
errors in interpreting data from the original sources.}
\label{tab:ZeroNuResults}
\begin{center}
\begin{tabular}{lcclc}  \hline\hline
Isotope               & Exposure            & Background  & Half-Life 
& \mee\    \\
                       & (kmole-y)           &  (counts)   & Limit 
(y)                  & (meV)    \\ \hline
        $^{48}$Ca      & $5\times 10^{-5}$   &   0         & $>1.4 
\times 10^{22}$       &  $<7200-44700^a$               \\
        $^{76}$Ge      &   0.467             &   21        & $>1.9 
\times 10^{25}$       &  $<350^b$              \\
        $^{76}$Ge      &   0.117             &   3.5       & $>1.6 
\times 10^{25}$       &  $<330-1350^c$               \\
        $^{76}$Ge      &   0.943             &   61        & $=1.2 
\times 10^{25}$       &  $=440^d$               \\
        $^{82}$Se      & $7\times 10^{-5}$   &   0         & $>2.7 
\times 10^{22}$(68\%) &  $<5000^e$               \\
        $^{100}$Mo     & $5\times 10^{-4}$   &   4         & $>5.5 
\times 10^{22}$       &  $<2100^f$              \\
        $^{116}$Cd     & $1\times 10^{-3}$   & 14          & $>1.7 
\times 10^{23}$       &  $<1700^g$              \\
        $^{128}$Te     & Geochem.            &   NA        & $>7.7 
\times 10^{24}$       &  $<1100-1500^h$              \\
        $^{130}$Te     & 0.025               &   5         & $>5.5 
\times 10^{23}$       &  $<370-1900^i$              \\
        $^{136}$Xe     & $7\times 10^{-3}$   &   16        & $>4.4 
\times 10^{23}$       &  $<1800-5200^j$              \\
        $^{150}$Nd     & $6\times 10^{-5}$   &   0         & $>1.2 
\times 10^{21}$       &  $<3000^k$              \\ \hline
\end{tabular}
\end{center}
$^a$Ogawa \etal  2004; $^b$Klapdor-Kleingrothaus \etal  2001; 
$^c$Aalseth \etal  2002; $^d$Klapdor-Kleingrothaus \etal  2004; 
$^e$Elliott \etal  1992;
$^f$Ejiri \etal  2001; $^g$Danevich \etal  2003; $^h$Bernatowicz \etal  
1993; $^i$Arnaboldi \etal  2004; $^j$ Luescher \etal  1998; $^k$De 
Silva \etal  1997
\end{table}

\subsubsection{A Claim for the Observation of \BBz}

In early 2002, a claim for the observation of \BBz\ was published 
(Klapdor-Kleingrothaus \etal  2002a).
The paper made a poor case for the claim and drew strong criticism 
(Aalseth \etal  2002a,
Feruglio \etal  2002, Zdesenko \etal  2002). The initial response to 
the criticism was emotional
(Klapdor-Kleingrothaus 2002b). In addition, one of the original 
co-authors wrote a
separate reply (Harney 2001) that mostly defended the claim yet 
acknowledged some significant
difficulty with the analysis. This author's name doesn't appear on 
later papers.
More recently, however, supporting evidence for the claim
has been presented and we recommend the reader study 
Klapdor-Kleingrothaus \etal (2002c) for
a good discussion of the initial evidence and Klapdor-Kleingrothaus 
\etal (2004) for the
most recent data analysis. Importantly, this later paper includes 
additional data and therefore an
increase in the statistics of the claim. In this subsection we 
summarize the current situation.
(We use the shorthand KDHK to refer to the collection of papers 
supporting the claim.)

\begin{figure}
\vspace{9pt}
\begin{center}
\epsfbox{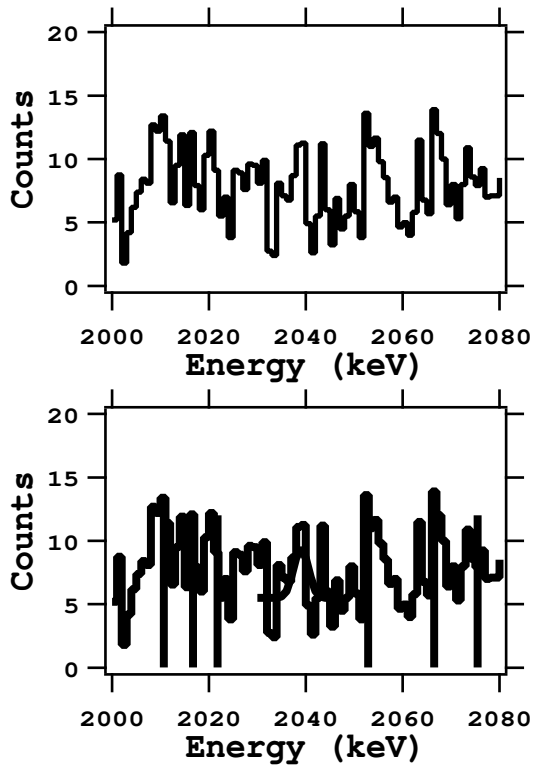}
\end{center}
\caption{The spectrum from the Heidelberg-Moscow experiment upon 
which the claim for \BBz\ is based.
The data in the two panels are identical. The lower panel has a 
Gaussian curve to indicate
the strength of the claimed \BBz\
peak.}
\label{fig:2}
\end{figure}

Figure \ref{fig:2} shows the spectrum corresponding to 71.7 kg-y 
of data from the Heidelberg-Moscow experiment
between 2000 and 2060 keV (Klapdor-Kleingrothaus \etal  2004).
This spectrum is shown here to assist the casual reader in understanding the
issues. However, the critical reader is encouraged to read the 
papers listed in the references as the
authors analyze several variations of this data using different techniques.
The fit about the
expected \BBz\ peak energy yields 28.75 $\pm$ 6.86 counts assigned 
to \BBz. The paper claims
a significance of approxmately $4\sigma$  for the peak, where the 
precise significance value depends on
the details of the analysis.
The corresponding best-fit lifetime, \Tz\ = 1.19 $\times$ 10$^{25}$ years
(Klapdor-Kleingrothaus \etal  2004), leads to a \mee\ of 440 meV 
with the matrix element
calculation of Staudt \etal (1990) chosen by the authors.

In the region between 2000 and 2100 keV, the KDHK analysis of 2002 
found a total of 7 peaks. Four of these
were attributed to $^{214}$Bi (2011, 2017, 2022, 2053 keV), one 
was attributed to \BBz\ decay (2039 keV),
and two were unidentified (2066 and 2075 keV). The KDHK analysis 
of 2004 does not discuss the spectrum
above 2060 in detail. An additional possible feature may also be present
near 2030 keV. A study (Klapdor-Kleingrothaus \etal  2003a) 
comparing simulation to calibration
with $^{214}$Bi demonstrates that if
the location of the Bi is known, the spectrum can be calculated. 
Furthermore, the relative strengths of
the strong Bi lines at 609, 1764 and 2204 keV can be used to 
determine the location of the activity.
Because the results of summing effects depend on the proximity of 
the activity,
its location is critical for the simulation of the weak peaks near 
the \BBz\ endpoint. The study also shows that the spectrum
can't be naively estimated, as was done in Aalseth \etal  (2002a). 
In fact, Table VII in Klapdor-Kleingrothaus
(2002c) finds, even with a careful simulation, that the expected 
strengths of the $^{214}$Bi peaks in the 2000-2100 keV
region are not predicted well by scaling to the strong peaks. That
is, the measured intensities of the weak peaks are difficult to
simulate without knowing the exact location of the activity. 
Furthermore, the deduced strengths of the weak lines
are more intense than  expected
by scaling from the strong peaks, even though the activity 
location is chosen to best
describe the relative intensities of the strong peaks.

Double-beta decay produces two electrons that have a short range 
in solid Ge. Therefore, the energy deposit
is inherently localized. Background process, such as the $\gamma$ 
rays from $^{214}$Bi, tend to produce
multiple energy deposits. The pulse waveform can be analyzed to 
distinguish single site events (SSE) from
multiple site events. Such an analysis by KDHK 
(Klapdor-Kleingrothaus \etal  2003c, 2004) tends to indicate that the 
Bi
lines and the unidentified lines behave as multiple site events, 
whereas the \BBz\ candidate events
behave as SSE. Note, however, that the statistics are still poor 
for the experimental lines and this conclusion
has a large uncertainty.
Nonetheless, this feature of the data is very intriguing and 
clearly a strength of the
KDHK analysis.

An analysis by Zdesenko \etal (2003) points out the strong 
dependence of the result on the choice of
the window width of the earlier 2002 analysis. The KDHK analysis 
argues that a small window is required because of the neighboring
background lines. Even so, their Monte Carlo analysis shows that 
the result becomes less stable for
small windows (see Fig. 9 in Klapdor-Kleingrothaus \etal  2002c). 
Zdesenko \etal (2003) also
remind us that the significance of a signal is overestimated when 
the regions used to estimate the
background are comparable to the region used to determine the 
signal (Narsky 2000). The report of
Klapdor-Kleingrothaus \etal (2004) fits a wide region containing 
several peaks simultaneously after
using a Bayesian procedure to identify the location of the peaks.

The claim for \BBz\ decay was made by a fraction of the 
Heidelberg-Moscow collaboration. A separate group of the
original collaboration presented their analysis of the data at the 
IV International
Conference on Non-Accelerator New Physics (Bakalyarov \etal  2003). 
They indicate that the data
can be separated into two distinct sets with different 
experimental conditions. One set includes
events that are described as ``underthreshold pulses'' and one set 
that does not. Analysis of the two
sets produce very different conclusions about the presence of the 
claimed peak. They conclude that the evidence
is an experimental artifact and not a result of \BBz decay.  KDHK responds 
(Klapdor-Kleingrothaus \etal  2004) that these corrupt data were not 
included in their analysis.

Traditionally, \BB\ experiments have ignored systematic 
uncertainties in their analysis. Only recently with
the start-up of high-statistics \BBt\ results has this situation 
begun to change. Historically, \BBz\ results
have always been quoted as upper limits based on low count rates. 
As a result, systematic uncertainties tended to
be negligible in the final quoted values. With a claim of a 
positive result, however, the stakes are dramatically
raised. It is clear that it is difficult to produce a convincing 
result when the signal counts are comparable
to expected statistical fluctuations in the background. The 
further presence of nearby unidentified peaks
makes the case even harder to prove. Although KDHK does discuss 
some systematic uncertainties qualitatively
and indicates they are small (in the position of the \BBz\ peak, 
and the expected peak width, for example),
there is no consideration of an uncertainty associated with the 
background model.

The next round of proposed \BBz\ experiments are designed to reach
$\sqrt{\delta m_{\rm atm}^{2}}$ and therefore will quickly
confirm or repudiate this claim. This is fortunate since the 
feature near 2039 keV in the KDHK claim
will likely require an experimental test. These experiments should 
provide a detailed listing of all
identified systematic uncertainties and a quantified estimate of 
their size. Furthermore, because the stakes
are very high and there will be many people who are biased, either 
for or against the KDHK claim, blind analyses
should also become part of the experimental design.

\subsection{Future Possibilities for \BBz Experiments}
The recent review by Elliott and Vogel (2002) describes the basics of 
experimental \BBz\ decay in some detail. Therefore,
we refer the reader to that article and only summarize the status of 
the various projects.
Table \ref{tab:ZeroNuProj} lists the proposals.

\begin{table}
\caption{\protect  A summary of the \BBz\ proposals. Background 
estimates were not available for all projects. The quantity
of isotope includes the estimated efficiency for \BBz.}
\label{tab:ZeroNuProj}
\begin{center}
\begin{tabular}{lrcl}  \hline\hline
Collaboration        & Isotope               &     Anticipated 
& Detector    \\
                      &    (kmol)             & Background 
& Description   \\
                      &                       &            (counts/y) 
&     \\ \hline
     CAMEO$^a$        &    $^{116}$Cd (2)     &      few/year 
& CdWO$_{4}$ crystals in liq. scint. \\
     CANDLES$^b$      &    $^{48}$Ca (0.04)   & 
& CaF$_{2}$ crystals in liq. scint.            \\
     COBRA$^c$        &                       & 
& CdTe semiconductors             \\
     CUORE$^d$        &    $^{130}$Te (1.4)   &    $\approx$ 60/y 
& TeO$_{2}$ bolometers           \\
     DCBA$^e$         &    $^{82}$Se (2)      &    $\approx$ 40/y 
& Nd foils and tracking chambers             \\
     EXO$^f$          &    $^{136}$Xe (4.2)   &    $<1$/y 
& Xe TPC,              \\
     GEM$^g$          &     $^{76}$Ge (11)    &    $\approx$ 0.8/y 
& Ge detectors in LN             \\
     GENIUS$^h$       &     $^{76}$Ge (8.8)   &    $\approx$ 0.6/y 
& Ge detectors in LN             \\
     GSO$^i$          &     $^{160}$Gd (1.7)  & 
& Gd$_{2}$SiO$_{5}$ crystals in liq. scint.             \\
     Majorana$^j$     &      $^{76}$Ge (3.5)  &    $\approx$ 1/y 
& Segmented Ge detectors            \\
     MOON$^k$         &      $^{100}$Mo (2.5) &    $\approx$ 8/y 
& Mo foils and plastic scint.             \\
     MPI bare Ge$^l$  &     $^{76}$Ge (8.8)   & 
& Ge detectors in LN             \\
     nano-crystals$^m$& $\approx$ 100 kmol     & 
& suspended nanoparticles             \\
     super-NEMO$^n$   &     $^{82}$Se (0.6)   &     $\approx$ 1/y 
& foils with tracking             \\
     Xe$^o$           &     $^{136}$Xe (6.3)  &     $\approx$ 118/y 
& Xe dissolved in liq. scint.             \\
     XMASS$^p$        &    $^{136}$Xe (6.1)   & 
& liquid Xe             \\  \hline
\end{tabular}
\end{center}
$^a$Bellini \etal  2001;
$^b$Kishimoto \etal  2004;
$^c$Zuber 2001;
$^d$Arnaboldi \etal  2004a;
$^e$Ishihara 2000;
$^f$Danilov \etal  2000;
$^g$Zdesenko \etal  2001;
$^h$Klapdor-Kleingrothaus \etal  2001;
$^i$Danevich \etal  2001, Wang \etal  2000;
$^j$Gaitskell \etal  2003;
$^k$Ejiri \etal  2000;
$^l$Abt \etal  2004;
$^m$McDonald 2004;
$^n$Sarazin \etal  2000;
$^o$Caccianiga and Giammarchi 2001;
$^p$Moriyama \etal  2001

\end{table}

\subsubsection{CANDLES}
The CANDLES collaboration has recently published the best limit on 
\BBz\ decay of $1.4 \times 10^{22}$ y
in $^{48}$Ca (Ogawa \etal  2004). Using the ELEGANTS VI detector, this
experiment consisted of 6.66 kg of CaF$_2$(Eu) crystals surrounded by 
CsI crystals, a layer of Cd, a layer of Pb,
a layer of Cu, and a layer of LiH-loaded paraffin, all enclosed 
within an air-tight box. This box
was then surrounded by boron-loaded water tanks and situated 
underground at the Oto Cosmo Observatory.
This measurement successfully demonstrated the use of these crystals 
for \BB\ studies.

An improved version of this crystal technology, the CANDLES-III 
detector (Kishimoto \etal  2004), is presently being constructed with 
200 kg
of CaF$_2$ crystals. These crystals have better light transmission 
than the CaF$_2$(Eu) crystals.
This design uses
sixty 10-cm$^3$ CaF$_2$ crystals, which are immersed in liquid
scintillator. The collaboration has
also proposed a 3.2-t experiment that hopes to reach 100 meV for \mee.

\subsubsection{COBRA}
The COBRA experiment (Zuber 2001) uses CdZnTe or CdTe semiconductor 
crystals. These crystals have many of the advantages of
Ge detectors but, in addition, operate at room temperature. Because 
the crystals contain Cd and Te, there are 7 \BB\ and
$\beta^+\beta^+$ isotopes contained. The final proposed configuration 
is for 64000 1-cm$^3$ crystals for
a total mass of 370 kg. The collaboration
has already obtained 30-keV resolution at 2.6 MeV with these 
detectors and has published initial \BB-decay studies (Kiel \etal  
2003). Background
studies are the current focus of the efforts. Although it is tempting 
to focus on the naturally isotopic abundant
$^{130}$Te for \BBz\ decay, the presence of the
higher Q-value $^{116}$Cd creates a serious background from its \BBt\ 
decay. Detectors enriched in $^{116}$Cd are
probably required to reach 45 meV.

\subsubsection{CUORE}
The CUORICINO experiment uses 41 kg of TeO$_2$ crystals operated at 
10 mK as bolometers. During the initial cool down,
some of the cabling failed and hence not all crystals were active. As 
a result the initial run had contained about
10 kg of $^{130}$Te (Arnaboldi \etal  2004). An initial exposure of 
5.46 kg-y, with an
energy resolution of 9.2 keV FWHM resulted in a limit \Tz\ $> 7.2 
\times 10^{23}$ y at 90\% confidence level (Norman 2004). The 
background in the region of interest for this
run was $0.22 \pm 0.04$ counts/(keV kg y). The experiment has been 
suspended in order to fix the cabling. Afterward, a
3-year run with the full mass will have a sensitivity of 10$^{25}$ years.

The CUORICINO project is a prototype for the CUORE proposal. CUORE 
would contain 760 kg of TeO$_2$. With the anticipated
improvement in background to better than 0.01 counts/(keV kg y), the 
half life sensitivity is $\approx 7 \times 10^{26}$ y or
a few 10's  of meV for \mee\ (Arnaboldi \etal  2004a).

\subsubsection{DCBA}
The Drift Chamber Beta-ray Analyzer (DCBA) (Ishihara 2000) is a 
tracking chamber within a 1.6-kG magnetic field that
can examine any \BB\ source that can be formed into a thin foil. On 
both sides of the foil are
tracking regions filled with 1-atm He gas. The future program calls 
for 25 m$^2$ of source foils contained
within each of 40 modules, for a total of approximately 600 kg of 
source material. The half-life
sensitivity is estimated to be a few $\times 10^{26}$ y for 
$^{82}$Se, $^{100}$Mo, or $^{150}$Nd, assuming
an enrichment of 90\%.

\subsubsection{EXO}
The EXO project proposes to use 1-10 t of about 80\% enriched 
liquid Xe as a time projection chamber
(Danilov \etal  2000). Development of a high-pressure gas TPC is being 
pursued in parallel.
In addition to measuring the energy deposit of the electrons, the 
collaboration is developing a technique
for extracting the daughter Ba ion from the Xe and detecting it 
offline. Observing the daughter in real time with the
\BB\ decay is a powerful technique for reducing background. With a 1-ton 
experiment, they anticipate
sensitivity to a lifetime of $8 \times 10^{26}$ y.

The collaboration has had some good progress on the research and 
development required to demonstrate that this technically
challenging project is feasible. They have determined the energy 
resolution by using both
ionization and scintillation measurements in liquid Xe.
The resolution result $\sigma$ = 3\% stated in Conti \etal (2003) was measured
at 570 keV. Assuming a statistical dependence on energy this means
about 1.5\% resolution at the \BBz\ energy of 2480 keV. They have 
also built an atom trapping system
and have observed lone Ba ions in an optical trap. Furthermore, they 
have begun experiments to demonstrate
that the ions are trapped and observable in an appreciable Xe gas 
background (Piepke 2004). Finally, using a
$^{222}$Ra source they are testing the Ba extraction technology. Ra 
and Ba have similar chemistry, but the
radioactive decay of Ra makes it a convenient test material.

The EXO team is currently preparing a 200-kg enriched-Xe experiment 
to operate at the Waste Isolation Pilot Plant (WIPP).
This prototype will not initially include Ba extraction.

\subsubsection{MOON}
The MOON project (Ejiri \etal  2000) proposes to use 1 t of Mo 
enriched to 85\% in $^{100}$Mo. The beauty of $^{100}$Mo is
that it not only is a good \BBz\ isotope, but also has a large 
charge-current cross section for low-energy
solar neutrinos. Thus the MOON detector is being designed to perform 
both experiments.
MOON measures individual $\beta$ rays from \BB\ decay, which helps 
identify events arising from the light-neutrino-mass
\BB\ mechanism and improve the background rejection.

The reference design for MOON is a collection of modules of 
interleaved plate and
fiber scintillators sandwiching Mo foils. Each foil is about 20 mg/cm$^2$.
Good position resolution is required to exploit the timing of the 
radioactive product
produced in the solar-neutrino interaction. The position is 
determined by the fiber scintillators,
whereas the scintillator plate provides
the energy resolution ($\sigma \approx 2.2$\% at 3 MeV).
Other detector options are under consideration but a 1-kg prototype 
of the reference design is currently being built.

\subsubsection{Majorana}
The Majorana Collaboration proposes to field 500 kg of 86\% enriched 
Ge detectors (Gaitskell \etal  2003).
By using segmented crystals and pulse-shape analysis, multiple-site 
events can be identified
and removed from the data stream. Internal backgrounds from 
cosmogenic radioactivities will be greatly
reduced by these cuts and external $\gamma$-ray backgrounds will also 
be preferentially eliminated.
Remaining will be single-site events like that due to \BB. The 
sensitivity is anticipated to be $4 \times
10^{27}$ y.

Several research and development activities are currently proceeding. 
The collaboration is building a multiple-Ge detector array, referred 
to as MEGA,
that will operate underground at the Waste Isolation Pilot Plant 
(WIPP) near Carlsbad, NM USA. This experiment will investigate the 
cryogenic cooling of many detectors sharing a cryostat in addition
to permitting studies of detector-to-detector coincidence techniques 
for background and signal identification.
A number of segmented crystals are also being studied to understand 
the impact of segmentation on
background and signal. This SEGA program consists of one 12-segment 
enriched detector and a number of commercially
available segmented detectors. Presently, commercially available 
segmented detectors are fabricated from
n-type crystals. Such crystals are much more prone to surface damage 
and thus more difficult to handle when
packaging inside their low-background cryostats. Hence the 
collaboration is also experimenting with segmenting p-type
detectors.

\subsubsection{Bare Ge Crystals}
The GENIUS collaboration (Klapdor-Kleingrothaus \etal  2001) proposed 
to install 1 t of enriched
bare Ge crystals in liquid nitrogen. By eliminating much of the 
support material surrounding
the crystals in previous experiments, this design is intended to 
reduce backgrounds of external origin.
Note how this differs from the background-reduction philosophy 
associated with pulse-shape analysis coupled
with crystal segmentation.
The primary advocates for this project indicate 
(Klapdor-Kleingrothaus \etal  2004)
that its motivation has
been questioned by their own claim of evidence for \BBz decay. Even 
so, the GENIUS test facility
(Klapdor-Kleingrothaus \etal  2003)
is being operated to demonstrate the effectiveness of operating 
crystals naked in liquid cryogen.

Another group at the Max Plank Institute in Heidelberg, however, is 
proposing to pursue a similar idea.
They have recently submitted a Letter of Intent (Abt 2004) to the 
Gran Sasso Laboratory. They propose
to collect the enriched Ge crystals from both the Heidelberg-Moscow 
and IGEX expeiments and
operate them in either liquid nitrogen or liquid argon. As a second 
phase of the proposal, they
plan to purchase an additional 20-kg of enriched Ge detectors (most 
likely segmented) and operate with a total of 35 kg
for about 3 years. Finally, they eventually plan to propose a large 
ton-scale experiment.
It should be noted that this collaboration and the Majorana 
collaboration are cooperating on
technical developments and if a future ton-scale experiment using 
$^{76}$Ge proceeds these
two groups will most likely merge and optimally combine the 
complementary technologies of bare-crystal operation
and PSA-segmentation.

\subsection{Nanocrystals}
Some elements may be suitable for
loading liquid scintillator with metallic-oxide nanoparticles. Since
Rayleigh scattering varies as the sixth power of the particle radius, it
can be made relatively small for nanoparticles of radii below 5 nanometers.
Particles of this size have been developed
and commercial suppliers of ZrO$_{2}$, Nd$_{2}$O$_{3}$ {\it etc}. are 
available.
Absorption of the materials must also be taken into account, but some of
the metal oxides such as ZrO$_{2}$ and TeO$_{2}$ are quite 
transparent in the optical
region because of the substantial
band gaps in these insulators. Some members of the SNO collaboration 
(McDonald 2004) have been studying
a configuration equivalent to filling the SNO cavity with a 1\% 
loaded liquid scintillator or approximately 10 t
of isotope after the present heavy water experiment is completed.
The group is currently researching the optical properties of
potential nano-crystal solutions. In particular, one must demonstrate 
that sufficient
energy resolution is achievable with liquid scintillator.

\subsubsection{Super-NEMO}
The recent progress of the NEMO-3 program (Sarazin \etal  2000) has 
culminated in excellent \BBt\ results.
In particular, the energy spectra from $^{100}$Mo
contain nearly 10$^5$ events and are nearly background free. These 
data permit, for the first
time, a precise study of the spectra. In fact, there is hope that 
the data (Sutton 2004) will demonstrate
whether the \BBt\ transition is primarily through a single 
intermediate state or through a
number of states (\v{S}imkovic \etal  2001).  The detector consists 
of several thin foils placed between
Geiger-drift cells, surrounded by a scintillator calorimeter.

NEMO-3 began operation in February 2003 with several isotopes, 
$^{100}$Mo being the most
massive at 7 kg, and plans to operate for 5 years. The 
collaboration plans to increase the mass of $^{82}$Se from 1 kg to 20 
kg and begin
an additional 5-year run. Presently, a Rn-removal trap is being 
installed to reduce the background, and
operation should begin again by the summer of 2004. The anticipated 
sensitivities for \Tz\ are $5 \times 10^{24}$y
and $3 \times 10^{25}$ yr for Mo and Se respectively. For the Se 
data, this corresponds to \mee\
below 100-200 meV.

A much bigger project is currently being planned that would use 100 
kg of source. The apparatus would have a large
footprint however and the Frejus tunnel where NEMO-3 is housed would 
not be large enough to contain it. Currently the collaboration
is studying the design of such a detector.

\subsubsection{XMASS}
The XMASS collaboration (Suzuki 2004; Moriyama 2001) plans to build a 
10 t natural Xe liquid scintillation
detector. They expect an energy resolution of 3\% at 1 MeV and hope 
to reach a value for \Tz\ $> 3.3 \times
10^{27}$ y. This detector would also be used for solar-neutrino 
studies and a search for dark matter.

\subsubsection{Borexino CTF}
In August of 2002, operations at the Borexino experiment resulted in 
the spill of scintillator. This
led to the temporary closure of Hall C in the Gran Sasso Laboratory 
and a significant change in operations at the underground
laboratory. As a result, efforts to convert the Counting Test 
Facility (CTF) or Borexino itself into a \BBz\ experiment
(Bellini 2001; Caccianiga and Giammarchi 2001) have been suspended 
(Giammarchi 2004).

\subsection{The Search for Decays to Excited States}
Searches for \BB\ to excited states in the daughter atom have been performed
in a number of isotopes but only observed in $^{100}$Mo. (The  experimental
situation is reviewed by Barabash 2000).
These experiments typically search for the $\gamma$ rays that 
characterize the excited states and
therefore are not mode-specific searches. The interpretation 
therefore is that the measured rate
(or limit) is for the \BBt\ mode. These data may be very useful to 
QRPA nuclear theory because
the behavior of the nuclear matrix elements with respect to $g_{pp}$ 
for the excited state decays is different
than for transitions to the ground state (Griffiths and Vogel 1992, 
Aunola and Suhonen 1996, Suhonen 1998). Thus, the excited state 
transitions probe different aspects
of the theory and may provide insight into the physics of the matrix elements.

A further reason for interest in decays to the excited state, as 
mentioned earlier, is the potential ability to discover
the process mediating the decay (\v{S}imkovic and Faessler 2002, 
Tomoda 2000). However, the decay rate to an excited state is 10-100 
times
smaller than rate to the ground state (Suhonen 2000a, 2000b). 
Furthermore the structure of the excited state in the daughter 
nucleus is
not as well understood as the ground state, and this increases the 
relative uncertainty in the nuclear matrix element.

\subsection{The Search for $\beta^+\beta^+$ Modes of Decay}
The $\beta^+$ modes of decay have not received the attention of the 
$\beta^-$ modes because of the
greatly reduced phase space and corresponding long half-lives. 
However, their detection would
provide additional matrix-element data. Furthermore, if the 
zero-neutrino mode were detected,
it might provide a handle on whether the decay is predominantly
mediated by a light neutrino or by right-handed currents (Hirsch \etal  1994).

Radiative neutrinoless double electron capture is a possible 
alternative to traditional neutrinoless
double beta decay (Sujkowski and Wycech 2003). In this process, two 
electrons are captured from
the atomic electron cloud and a radiated photon carries the full 
Q-value for the decay. A resonance condition can enhance the rate 
when the energy release
is close to the 2P-1S energy difference. In this case, high-Z, 
low-energy-release isotopes are favored (e.g. $^{112}$Sn).
Unfortunately the mass differences for the candidate isotopes are not 
known precisely enough to
accurately predict the overlap between the two energies. If a 
favorable overlap does exist, however, the
sensitivity to \mee\ might rival that of \BBz decay.

\subsection{Towards a 100-kg experiment}
The KDHK spectrum shows a feature very close to the \BBz\ endpoint. 
This intriguing result will need to be
confirmed or refuted experimentally. One can see the required 
operation parameters for a confirmation experiment
from Klapdor-Kleingrothaus \etal (2004). One needs about 75 kg-y 
of exposure,
and a background lower than about 0.5 counts/(kg y).  Note that 
most of the proposals described above will all accomplish this very 
early on in their program if they meet their design goals.
If instead one designs an experiment only to test the claim (not to 
provide a precise measurement of the
\Tz) then a 100-kg experiment could provide the answer after a modest run time.

If the KDHK result holds up, it will be a very exciting time for 
neutrino-mass research. A \mee\ near 400 meV
means that $\beta$-decay experiments and cosmology will be sensitive 
to the mass. As a result, one can certainly
imagine a not-too-distant future in which we know the neutrino mass 
and its Majorana-Dirac character. Towards
this goal, a precision measurement of \mee\ will be required. To 
accomplish this, we will need more than
one \BB\ experiment, each with a half-life measurement accurate to 10-20\%.
At this level the uncertainty will
be dominated by the matrix element uncertainty even if future 
calculations can be trusted to 50\%.
With two experiments utilizing different isotopes, one might 
disentangle the uncertainty in \Mz.

\subsection{Towards the 100-ton experiment}
The next generation of experiments hopes to be sensitive to 
$\sqrt{\delta m_{\rm
atm}^{2}}$. If they fail
to see \BBz\ at that level, the target for the succeeding 
generation of efforts will be
$\sqrt{\delta m_{\rm sol}^{2}}$. This scale is an order of magnitude 
lower and hence will require two
orders of magnitude more isotopic mass, 
approximately 100 tons of isotope.

A 100-ton experiment will have to face the same technical challenges 
associated with radioactive backgrounds
and energy resolution as today's proposals. (See Elliott and Vogel 
2002 for a discussion of these issues.)
In addition, a background from solar neutrinos will also
have to be considered. Solar neutrinos can result in a background via 
elastic scattering or charged current
interactions. The rate ($R_{\beta\beta}$) of \BBz\ events can be written

\begin{equation}
\label{eqn:BBRate}
R_{\beta\beta} = \frac{1}{M}\frac{dN}{dt} = \frac{\lambda N}{M} 
\approx \frac{420}{MW(g)}\left(\frac{10^{27}y}{T^{0\nu}_{1/2}} 
\right) {\rm y^{-1}  t^{-1}}~,
\end{equation}
where $MW(g)$ is the molecular weight of the \BB\ isotope and
$M$ is the mass of the target in tons. 
For pure $^{136}$Xe, this would result in about 3
\BBz\ events per year per ton for \Tz\ = 10$^{27}$ y.

Elastic scattering (ES) rates from $^8$B solar neutrinos can be 
comparable to this \BBz\ rate if the
target material contains the isotope at a low fraction. For example, 
a 2\% solution of Xe in liquid scintillator
has been discussed as a possible \BBz\ experiment.  Of course the 
number of ES events within the \BBz\ window depends on the resolution 
and therefore
we need the cross section per unit energy ($\Delta \sigma/\Delta 
E$)(Bahcall 1989)
\begin{equation}
\label{eqn:ESCross}
\frac{\Delta \sigma}{\Delta E} \approx 9 \times 10^{-48} {\rm cm}^2/{\rm keV}~.
\end{equation}
The rate of ES events ($R_8$) is then given by
\begin{eqnarray}
\label{eqn:B8Rate}
R_8 & = & (F_8 \frac{\Delta \sigma}{\Delta E} N_A)\left(\Delta E 
\frac{1}{MW_t(g)}N_e\right)   \nonumber \\
     & \approx & (8 \times 10^{-4} /({\rm keV \ y \ t}))\left(\Delta E 
\frac{M}{MW_t(g)}N_e\right)
\end{eqnarray}
where $F_8$ is the $^8$B neutrino flux ($5 \times 10^{6}$ 
/(cm$^2$ s)), $N_A$ is Avagadro's number, $\Delta E$ is the
\BBz\ energy window in keV, $MW_t$ is 
the target molecular weight, and $N_e$ is the
number of electrons per molecule of the target.  For a pure Xe target with
an energy window of 50 keV, we find $R_8 \approx 0.02/({\rm y \ t})$.  This 
background is not a problem for any pure Xe detector that proposes 
a half-life sensitivity of $10^{28}$ y. It is  significant for a detector 
with only 2\% Xe at \Tz=$10^{27}$ y.

Charge-current ($CC$) scattering of solar neutrinos, especially
the large flux of solar pp neutrinos may also be a background for \BBz\ decay.
As pointed out by Raghavan (1997), some \BB\ isotopes make 
interesting targets for pp solar neutrino experiments
because the reaction produces a radioactive isotope, the intermediate 
nucleus in the \BB\ process. The decay of this
product nucleus provides a coincidence signature for the $CC$ 
reaction. However for several reasons, this process
must be considered as a background for \BBz\ decay. First, the 
$\beta$-decay Q-value of the intermediate
nucleus is larger than the \BB\ Q-value for most of the nuclei. 
Furthermore, the half-life of the intermediate
nucleus is often too long for an effective coincidence 
identification. Finally, the end product $\beta$-decay daughter
nucleus is the same as the \BB\ daughter.

The  rate
of $CC$ events ($R_{CC}$) can be written:

\begin{equation}
\label{eqn:CCRate}
R_{CC} = F_{CC} \ \langle \sigma_{CC} \rangle \frac{N_A}{MW(g)} f_E = 
18 ~  R_{SNU} \frac{f_E}{MW(g)} (y^{-1} t^{-1})~,
\end{equation}
where $F_{CC}$ is the solar neutrino flux, $\langle 
\sigma_{CC} \rangle$ is the spectrum-weighted $CC$ cross section, 
$f_E$ is
the fraction of intermediate nucleus $\beta$ decays that fall within 
the \BBz\ energy window, and $R_{SNU}$
is the rate of $CC$ interactions in SNU ($10^{-36}$ interactions/(s 
target atom)).
For example, the total $R_{CC} \approx 120/({\rm y \ t})$ (that is with 
$f_E=1$) in the MOON proposal
is much higher than 
$R_{\beta\beta} \approx 4/(\rm{y \ t})$ for a half-life of $10^{27}$ y.
In that case however, the intermediate nucleus has a convenient 
lifetime of 16 s, short enough
that a delayed coincidence can identify the decay and separate it 
event-by-event from \BBz\ candidate events.

$R_{CC}$ in each \BB\ isotope must be considered separately. 
For example, $^{76}$Ge, $^{116}$Cd, and $^{130}$Te
do not have a low-lying intermediate
nucleus level reachable by the high-flux pp neutrinos, and in the case 
of Ge, $^{7}$Be neutrinos. Alternatively, $^{82}$Se, $^{176}$Yb, and 
$^{160}$Gd
have significant cross sections
but the long-lived intermediate nucleus prevents easy 
anti-coincidence identification.
In the important case of $^{136}$Xe, little is known about the
structure of the intermediate-nucleus ($^{136}$Cs) excited states. 
The decay of $^{136}$Cs is to highly excited levels
in $^{136}$Ba that decay themselves via $\gamma$ emission. This results in a 
relatively large $f_E$, but also a $\gamma$-ray
cascade that might provide a signature to eliminate the potential 
background. $^{150}$Nd is another isotope in which little is known about 
the intermediate-nucleus excited states.

\section{Conclusion}
If the reader retains anything from this review, it should be that 
the information
recently acquired from oscillation experiments makes
this an exciting time for \BB\ decay because the next generation of experiments
will be sensitive to neutrino masses on the order of $\sqrt{\delta 
m_{\rm atm}^{2}}.$
If a nonzero rate is seen, we will know that neutrinos are Majorana particles.
With some progress on the calculation of nuclear matrix elements --- 
and we believe
progress is possible --- a nonzero rate in these experiments should allow the
determination of
the hierarchy realized by nature and the absolute mass scale.  If no 
signal is seen,
we should be able to say either that the hierarchy is normal or that 
neutrinos are Dirac particles.

Although none of the next-generation experiments is ready to operate, 
many of the
new proposals are promising. We are hopeful that with concentrated effort from
theorists and experimentalists, \BB\ decay will add to our growing 
understanding of neutrinos.

\ack
   Many people helped us in preparing this manuscript. In particular, Frank
Avignone and Petr Vogel read it critically.

A number of our experimental colleagues assisted by
ensuring that our descriptions of their work were accurate. We are indebted
to Ron Brodzinski,
Hiroyasu Ejiri, Marco Giammarchi, Werner Hofman, Nobuhiro Ishihara, 
Tadafumi Kishimoto,
Arthur McDonald, Rick Norman, Andreas Piepke, Yoichiro Suzuki, and Kai Zuber.

   We would like to thank Victor Gehman for his gracious assistance with Fig.
\ref{fig:masses3d}, and Chris Kolda and Jouni Suhonen  for useful discussion.

   This work was supported in part by the U.S. Department of Energy
under grant DE-FG02-97ER41019 and by LANL
Laboratory-Directed Research and Development.

\section*{References}
%
%


\begin{harvard}

\item[] Aalseth C E \etal  2002 {\it Phys. Rev.} D {\bf 65} 092007
\item[] Aalseth C E \etal  2002a {\it Mod. Phys. Lett.} A {\bf 17} 1475-1478
\item[] Abada A and Bhattacharyya G 2003 {\it Phys. Rev.} D {\bf 68} 033004
\item[] Abt I \etal  2004 LNGS-LOI 35/04 
\item[] Ahmed S N \etal 2004  {\it Phys. Rev. Lett.} {\bf 92} 181301
\item[] Allen S W \etal  2003 {\it Mon. Not. Roy. Astron. Soc.} {\bf 346} 593
\item[] Arnaboldi C \etal  2004 {\it Phys. Lett.} B {\bf 584} 260-268 
\item[] Arnaboldi C \etal  2004a {\it Nucl. Instrum. Meth.} {\bf 
A518} 775 
\item[] Aunola M and Suhonen J 1996 {\it Nucl. Phys.} A {\bf 602} 133
\item[] Aunola M and Suhonen J 1998 {\it Nucl. Phys.} A {\bf 643} 207
\item[] Avignone F T III 2004 private communication to Steve Elliott
\item[] Bahcall John N and Pe\~{n}a-Garay C 2003 {\it JHEP} {\bf 0311} 004
\item[] Bahcall J N 1989 { \it Neutrino Astrophysics} (Cambridge:
Cambridge University Press)
\item[] Bakalyarov A M, Balysh A Ya, Belyaev S T, Lebedev V I and 
Shukov S V 2003 {\it Preprint} hep-ex/0309016
\item[] Barabash A S 2000 {\it Czhech. J. Phys.} {\bf 50} 447 
\item[] Barbero C, Krmpotic F, Mariano A and Tadic D 1999 {\it 
Nucl. Phys.} A {\bf 650} 485
\item[] Barger V, Glashow S L, Langacker P and Marfatia D 2002 {\it 
Phys. Lett.} B {\bf 540} 247
\item[] Barger V \etal  2003 {\it Preprint} hep-ph/0312065
\item[] Bellini G \etal  2001 {\it Eur. Phys. J.} C {\bf 19} 43  
\item[] Bender M, Dobaczewski J, Engel J and Nazarewicz W, 2002 {\it 
Phys. Rev.} C {\bf 65} 054322
\item[] Bennett C L \etal  2003 {\it Astrophys. J. Suppl.} {\bf 148} 1  
\item[] Bernatowicz T \etal  1993 {\it Phys. Rev.} C {\bf 47} 806
\item[] Bilenky S M \etal  1999 {\it Phys. Lett. B} {\bf 465} 193
\item[] Bilenky S M \etal  2003 {\it Phys.Rept.} 379  69
\item[] Bilenky S M, Faessler A and \v{S}imkovic F 2004 {\it Preprint}
hep-ph/0402250
\item[] Bobyk A, Kaminski W A and \v{S}imkovic F 2001 {Phys. Rev.} {\bf 
C63} 051301(R)
\item[] Bornschein L 2003 {\it Preprint} hep-ex/0309007 
\item[] Caccianiga B and Giammarchi M G 2001 {\it Astropart. Phys.} 
{\bf 14} 15 
\item[] Carlson J 1998 {\it Rev. Mod. Phys.} {\bf 70} 743
\item[] Caurier E, Nowacki F, Poves A and Retamosa J 1996 {\it Phys. 
Rev. Lett.} {\bf 77} 1954
\item[] Civitarese O and Suhonen J 2003 {\it Nucl. Phys.} A {\bf 729} 867
\item[] Conti E \etal  2003 {\it Phys. Rev.} B {\bf 68} 054201 
\item[] Croft R A C \etal  2002 {\it Astrophys. J.} {\bf 581} 20 %
lyman alpha forest
\item[] Crotty P, Lesgourgues J and Pastor S 2004 {\it Preprint} hep-ph/0402049
\item[] Czakon M \etal  {\it Phys. Rev. D} to be published
\item[] Danilov M \etal  2000 {\it Phys. Rev.} C {\bf 62} 044501 
\item[] Danevich F A \etal  2001 {\it Nucl. Phys.} A {\bf 694} 375 
\item[] Danevich F A \etal  2003 {\it Phys. Rev.} C {\bf 68} 035501
\item[] de Holanda P C and Smirnov A Yu 2003 {\it Preprint} hep-ph/0309299
\item[] De Silva A \etal  1997 {\it Phys. Rev.} C {\bf 56} 2451
\item[] Doi M, Kotani T and Takasuga E 1985 {\it Prog. Theor. Phys. 
Suppl.} {\bf 83} 1
\item[] Dussel G G, Maqueda E and Perazzo R P J 1970 {\it Nucl. Phys.} A {\bf
153} 469
\item[] Ejiri H \etal  2001 {\it Phys. Rev.} C {\bf 63} 065501
\item[] Ejiri H \etal  2000 {Phys. Rev. Lett.} {\bf 85} 2917 
\item[] Elgar{\o}y {\O} \etal  2002 {\it Phys. Rev. Lett.} {\bf 89} 
061301  
\item[] Elgar{\o}y {\O} and Lahav O 2003 {\it JCAP} {\bf 0304} 004
\item[] Elliott S R \etal  1992 {\it Phys. Rev.} C {\bf 46} 1535
\item[] Elliott S and Vogel P 2002 {\it Annu. rev. Nucl. Part. Sci.} 
{\bf 52} 115-51
\item[] Engel J, Bender M, Dobaczewski J, Nazarewicz W and  Surman 
R 1999 {\it Phys. Rev.} C {\bf 60} 14302
\item[] Engel J, Pittel S, Stoitsov M, Vogel P and Dukelsky J 1997 
{\it Phys. Rev.} C {\bf 55}1781
\item[] Engel J, Vogel P and Zirnbauer M R 1988 {\it Phys. Rev.} 
C {\bf 37} 731
\item[] Engel J and Vogel P 2004 {\it Phys. Rev.} C {\bf 69} 034304
\item[] Engel J, Vogel P, Ji X and Pittel S 1989 {\it Phys. Lett.} 
B {\bf 225} 5
\item[] Evans J A, Dussel G G, Maqueda E E and Per-azzo R P J 1981 
{\it Nucl. Phys.} A {\bf 367} 77; Dussel G G, Maqueda E E, Perazzo R P J and
Evans J A 1986 
{\it Nucl. Phys.} A {\bf 450} 164
\item[] Faessler A, Kovalenko S, \v{S}imkovic F and Schwieger J 1997 {\it
Phys. Rev. Lett.} {\bf 78} 183
\item[] Faessler A and \v{S}imkovic F 1998 {\it J. Phys. G: Nucl. 
Part. Phys.} {\bf 24} 2139
\item[] Feruglio F, Strumia A and Vissani F 2002 {\it Nucl. Phys.} 
B {\bf 637} 345
\item[] Fukugita M and Yanagida T 1986 {\it phys. Lett.} B {\bf 174} 45
\item[] Gaitskell R \etal  2003 {\it Preprint} nucl-ex/0311013 
\item[] Gell-Mann M \etal  1979 in \textit{Supergravity} (Amsterdam: 
North Holland) p~315
\item[] Giammarchi M G 2004 private communication 
\item[] Ginocchio J N 1980 {\it Ann. Phys.} {\bf 126} 234
\item[] Giunti C 2003 {\it Preprint} hep-ph/0308206
\item[] Griffiths A and Vogel P 1992 {\it Phys. Rev.} C {\bf 46} 181
\item[] Hableib J A and Sorensen R A 1967 {\it Nucl. Phys.} A {\bf 98} 542
\item[] Hagiwara K \etal  2002 {\it Phys. Rev. D} {\bf 66} 1
\item[] Hannestad S 2003 {\it Preprint} hep-ph/0312122 
\item[] Hannestad S 2003a {\it JCAP} {\bf 0305} 004
\item[] Hara K 1964 {\it Prog. Theor. Phys.} {\bf 32} 88
\item[] Harney H L 2001 {\it Mod. Phys. Lett.} A {\bf 16} 2409
\item[] Haxton W C and Song C-L 2000 {\it Phys. Rev. Lett.} {\bf 84} 5484
\item[] Haxton W C and Stephenson G F 1984 {\it Jr. Prog. Part. 
Nucl. Phys.} {\bf 12} 409
\item[] Hecht K T 1993 {\it J. Phys.} A {\bf 26} 329
\item[] Hirsch M \etal  1994 {\it Z. Phys.} A {\bf 347} 151 
\item[] Hirsch J G, Castanos O, Hess P O and Civitarese O 2002 {\it 
Phys. Lett.} B {\bf 534} 57
\item[] Hirsch J, Castanos O, Hess P O and Civitarese O 2002 {\it 
Phys. Rev.} C {\bf 66} 015502
\item[] Hirsch J, Hess P O and Civitarese O 1997 {\it Phys. Rev.} 
C {\bf 56} 199
\item[] Hu W and Tegmark M 1999 {\it Ap. J. Lett.} {\bf 514} L65
\item[] Huffman A H 1970 {\it Phys. Rev.} C {\bf 2} 742
\item[] Ishihara N \etal  2000 {\it Nucl. Instrum. Meth.} A {\bf 443} 101 
\item[] Ishitsuka M 2004 Presented at ``The 5$^{th}$ Workshop on 
Neutrino Oscillations and their Origins (NOON2004),
Tokyo, Japan, Feb. 10-15 
\item[] Joaquim F R 2003 {\it Phys. Rev.} D {\bf 68} 033019
\item[] Kaplan D B, Nelson A E and Weiner N 2004 {\it Preprint} hep-ph/0401099
\item[] Ke Y \etal  1991 {\it Phys. Lett.} B {\bf 265} 53
\item[] Kiel H, M\"{u}nstermann D and Zuber K 2003 {\it Preprint}
nucl-ex/0301007 
\item[] Kishimoto T {\it et al.} 2000 {\it Osaka University 
Laboratory for Nuclear Studies Annual Report}; Kishimoto T 2004, 
private communication 
\item[] Klapdor-Kleingrothaus H V 2000 {\it Springer Tracts in Mod. 
Phys.} {\bf 163} 69
\item[] Klapdor-Kleingrothaus \etal  2001 {\it Eur. Phys. J.} A {\bf 12} 147
\item[] Klapdor-Kleingrothaus \etal  2001 {\it Workshop on 
``Neutrino Oscillations and Their Origin", NOON'2000} (Singapore: 
World Scientific) 
\item[] Klapdor-Kleingrothaus \etal  2003 {\it Nucl. Instrum. Meth.} 
A {\bf 511} 341-346 
\item[] Klapdor-Kleingrothaus H V, P\"{a}s H and Sarkar U 1999 {\it 
Eur. Phys. J.} A {\bf 5} 3 
\item[] Klapdor-Kleingrothaus H V, P\"{a}s H and Smirnov A Yu 2001 
{\it Phys. Rev. D} {\bf 63} 073005
\item[] Klapdor-Kleingrothaus H V, Dietz A, Harney H L and 
Krivosheina I V 2002a {\it Mod. Phys. Lett.} A {\bf 16} 2409-2420
\item[] Klapdor-Kleingrothaus H V 2002b {\it Preprint} hep-ph/0205228
\item[] Klapdor-Kleingrothaus H V, Dietz A and Krivosheina I V 
2002c {\it Found. Phys.} {\bf 32} 1181-1223
\item[] Klapdor-Kleingrothaus H V, Dietz A, Krivosheina I V and  O. 
Chkvorets 2004 {\it Phys. Lett.} B {\bf 586} 198;
   {\it Nucl. Instrum. Meth.} A {\bf 522} 371
\item[] Klapdor-Kleingrothaus H V, Chkvorez O, Krivosheina I V and Tomei 
C 2003a {\it Nucl. Instrum. Meth.} A {\bf 511} 355
\item[] Kobzarev I Yu \etal  1980 {\it Sov. J. Nucl. Phys.} {\bf 32} 823
\item[] Kowalski K, Dean D J, Hjorth-Jensen M, Papenbrock T and 
Pieuch P 2003 {\it Phys. Rev. Lett.} {\bf 92} 132501
\item[] Kuo C \etal  2004 {\it Astrophys. J.} {\bf 600} 32
\item[] Lobashev V M \etal  1999 {\it Phys. Lett.} B {\bf 460} 227  
\item[] Luescher R \etal  1998 {\it Phys. Lett.} B {\bf 434} 407
\item[] Matsuda K \etal  2001 {\it Phys. Rev. D} {\bf 64} 013001
\item[] McDonald A 2004 private communication for members of the 
SNO Collaboration 
\item[] Miller G A and Spencer J E 1976 {\it Ann. Phys.}{\bf 100} 562
\item[] Moe M and Vogel P 1994 {\it Ann. Rev. Nucl. Part. Sci} {\bf 44} 247
\item[] Mohapatra R and Senjanovic G 1980 {\it Phys. Rev. Lett.} {\bf 44} 912
\item[] Moriyama S \etal  2001 presented at XENON01 Workshop, Dec., Tokyo, 
Japan
\item[] Murayama H and Pe\~{n}a-Garay C 2004 {\it Phys. Rev.} D {\bf 69} 031301
\item[] Muto K, Bender E and Klapdor-Kleingrothaus H V 1989 {\it Z. 
phys.} A {\bf 334} 187
\item[] Narsky I 2000 {\it Nucl. Instrum. Meth.} A {\bf 450} 444
\item[] Norman E 2004 private communication 
\item[] Ogawa I \etal  2004 {\it Nucl. Phys.} A {\bf 730} 215 
\item[] Osipowicz A \etal  2001 {\it Preprint} hep-ex/0109033 
\item[] Pantis G, \v{S}imkovic F, Vergados J D and Faessler A 1999 
{\it Phys. Rev.} C {\bf 60} 055502
\item[] Papenbrock T and Dean D J 2003 {\it Phys. Rev.} C {\bf 67} 051303
\item[] Papenbrock T, Juodagalvis A and Dean D J 2003 {\it Phys. Rev.} {\bf
C69} 024312
\item[] Pascoli S, Petcov S T and Wolfenstein L 2002 {\it Phys. 
Lett.} B {\bf 524} 319
\item[] Pascoli S, Petcov S T and Rodejohann W 2002a {\it Phys. Lett.} 
{\bf 549} 177
\item[] Pascoli S and Petcov S T 2003 {\it Proc. Xth 
Int. Workshop on
          Neutrino Telescopes} March 11-14, Venice, Italy
\item[] Pascoli S and Petcov S T 2004 {\it Phys. Lett.} B {\bf 580} 280; 
{\it Phys. Lett.} B {\bf 544} 239
\item[] Person T J \etal  2002 {\it Preprint} astro-ph/0205388    
\item[] Piepke A 2004 private communication
\item[] Planck 2003 See http://astro.estec.esa.nl/Planck 
\item[] Primakoff H and Rosen SP 1959 {\it Rep. Progr. Phys.} {\bf 22} 121
\item[] Raduta A A, Faessler A and Stoica S 1991 {\it Nucl. Phys.} 
A {\bf 534} 149
\item[] Raghavan R 1997 presented at {\it Fourth Int. Solar 
Neutrino Conf.} Heidelberg, Germany,
           (Max-Planck-Institut f\"{u}r Kernphysik Heidelberg) p~249
\item[] Rodin V A, Faessler A, \v{S}imkovic F and Vogel P 2003 {\it 
Phys. Rev.} C {\bf 68} 044302
\item[] Rowe D J 1968 {\it Rev. Mod. Phys.} {\bf 40} 153
\item[] Prezeau G, Ramsey-Musolf M and Vogel P 2003 {\it Phys. Rev.} C {\bf
68} 034016
\item[] Saji C, 2004 Presented at ``The 5$^{th}$ Workshop on 
Neutrino Oscillations and their Origins (NOON2004),
Tokyo, Japan, Feb. 10-15 
\item[] Sarazin X \etal  2000 {\it Preprint} hep-ex/0006031
\item[] Schechter J and Valle J W F 1982 {\it Phys. Rev.} D {\bf 25} 25091
\item[] Schwieger J, \v{S}imkovic F and Faessler A 1996 {\it Nucl. 
Phys.} A {\bf 600} 179-192
\item[] SDSS 2003 See http://www.sdss.org  
\item[] Siiskonen T, Hjorth-Jensen M and Suhonen J 2001 {\it Phys. 
Rev.} C {\bf 63} 055501
\item[] \v{S}imkovic F and Faessler A 2002 {\it Prog. Part. Nucl. 
Phys.} {\bf 48} 201
\item[] \v{S}imkovic F, Pacearescu L and Faessler A 2003 {\it Nucl. Phys.} A
{\bf 733} 321 
\item[] \v{S}imkovic F, Pantis G, Vergados J D and Faessler A 1999 
{\it Phys.  Rev.} C {\bf 60} 055502
\item[] \v{S}imkovic F, Schwinger S, Veselsky M, Pantis G and 
Faessler A 1997 {\it Phys. Lett.} B {\bf 393} 267
\item[] \v{S}imkovic F \etal  2001 {\it J Phys. G: Nucl. Part. 
Phys.} {\bf 27} 2233-2240
\item[] Smy M \etal  2003 {\it Preprint} hep-ex/0310064
\item[] Spergel D N \etal  2003 {\it Astrophys. J. Suppl.} {\bf 148} 175
\item[] Staudt A, Muto K, and Klapdor H V 1990 {\it Europhys. Lett.} 
{\bf 13} 31
\item[] Stoica S and Klapdor-Kleingrothaus H V 2001 {\it Nucl. 
Phys.} A {\bf 694} 269
\item[] Sugiyama H 2003 {\it Preprint} hep-ph/0307311
\item[] Suhonen J 1998 {\it Nucl. Phys. of the Russ. Acad. Yad. 
Fiz.} {\bf 61} 1286
\item[] Suhonen J 2000a {\it Phys. Lett.} B {\bf 477} 99-106
\item[] Suhonen J 2000b {\it Phys. Rev.} C {\bf 62} 042501(R)
\item[] Suhonen J and Civitarese O 1998 {\it Phys. Rep.} {\bf 300} 123
\item[] Suhonen J, Civitarese O and Faessler A 1992 {\it Nucl. Phys.} 
A {\bf 543} 645
\item[] Suhonen J, Divari P C, Skouras L D and Johnstone I P 1997 
{\it Phys. Rev.} C {\bf 55} 714
\item[] Sujkowski Z and Wycech S 2003 {\it Preprint} hep-ph/0312040; Wycech S
and 
Sujkowski Z 2004 {\it Acta. Phys. Polon.} B {\bf 35} 1223  
double EC
\item[] Sutton S 2004 private communications
\item[] Suzuki Y 2004 private communication 
\item[] Tegmark M \etal  2003 {\it Preprint} astro-ph/0310723
\item[] Tomoda T, Faessler F, Schmid K W and Grummer F, 1986 {\it Nucl. 
Phys.} {\bf  A452} 591
\item[] Toivanen J and Suhonen J 1995 {\it Phys. Rev. Lett} {\bf 75} 410-413
\item[] Tomoda T 1991 {\it Rep. Prog. Phys.} {\bf 54} (1991) 53
\item[] Tomoda T 2000 {\it Phys. Lett.} B {\bf 474} 245-250
\item[] Tomoda T and Faessler A 1987 {\it Phys. Lett.} B {\bf 199} 475
\item[] Vergados J D 2000 {\it Phys. Atom. Nucl.} {\bf 63} 1137
\item[] Vissani F 1999 {\it JHEP} {\bf 06} 022
\item[] Vogel P and Zirnbauer M R 1986 {\it Phys. Rev. Lett.} {\bf 57} 3148
\item[] Wang S C, Wong H T and Fujiwara M 2000 {\it Nucl. Instrum. 
Meth.} A {\bf 479} 498-510 
\item[] Weinheimer C \etal  1999 {\it Phys. Lett.} B {\bf 460} 219  
\item[] Wu C-L, Feng D H and Guidry M W 1994 {\it Adv. Nucl. Phys.} 
{\bf 21} 227
\item[] Xing Z 2003 {\it Phys. Rev.} D {\bf 68} 053002
\item[] Yanagida T 1979  {\it Proc. Workshop on Unified Thery
and Baryon Number in the Universe} (KEK,
Tsuuba, Japan)
\item[] Zdesenko Yu G, Ponkratenko O A and Tretyak V I 2001 {\it J. 
Phys.} {\bf G27} 2129 
\item[] Zdesenko Yu G, Danevich F A and Tretyak V I 2002 {\it Phys. 
Lett.} B {\bf 546} 206-215
\item[] Zuber K 2001 {\it Phys. Lett.} B {\bf 519} 1  

\end{harvard}

\end{document}